\documentclass[10pt]{article}
\usepackage{graphicx}
 
\usepackage{epsfig}
\usepackage{latexsym}
\usepackage{amsmath}

\begin{document}

\title{Slow-down or speed-up of inter- and intra-cluster diffusion of controversial knowledge in stubborn communities based on  a small world    network }

\author{ M. Ausloos$^{1,2,3,}$\footnote{ 
$e$-$mail$ $address$:
marcel.ausloos@ulg.ac.be;  ma683@le.ac.uk}
   \\ 
$^1$
School of Management, University of Leicester,\\
 University Road,
Leicester, LE1 7RH, UK\\
$^2$
e-Humanities group\footnote{Associate Researcher}, Royal Netherlands Academy (NKV), \\Joan Muyskenweg 25, 1096 CJ
Amsterdam, The Netherlands\\
$^3$
GRAPES, Beauvallon Res., rue de la Belle Jardiniere, 483/0021\\
B-4031, Liege Angleur, Euroland  \\
 }
\date{\today}
\maketitle
\vskip -.5truecm

\begin{abstract}
  Diffusion of knowledge is expected to be huge when agents are open minded. The report concerns a more difficult diffusion case when communities are  made of stubborn agents.     Communities having markedly different opinions are for example the  Neocreationist and Intelligent Design  Proponents (IDP), on one hand, and the Darwinian  Evolution Defenders (DED),  on the other hand. The case of knowledge diffusion within such communities is studied here  on a network based on an adjacency matrix  built from time ordered selected quotations of agents, whence for inter- and intra-communities. The network is intrinsically directed and not necessarily reciprocal.  Thus, the adjacency matrices have complex eigenvalues;  the eigenvectors present complex components. A  quantification of   the slow-down or speed-up  effects of information diffusion in  such temporal networks, with non-Markovian contact sequences, can be made by comparing the real time dependent (directed) network to its counterpart, the time aggregated (undirected) network, - which has real eigenvalues. In order to do so,  
  small world  networks which both contain an $odd$ number of nodes are studied and compared to similar  networks with an $even$ number of nodes. 
    It is found that  (i)  the diffusion of knowledge is more difficult on the largest networks;  (ii)  the network size influences the slowing-down or speeding-up diffusion process. Interestingly, it is observed that  (iii) the  diffusion of knowledge is slower in IDP and  faster in DED communities. It is suggested that the finding can be "rationalized", if   some "scientific quality"  and "publication habit" is attributed to the agents, as common sense would guess.   This  finding offers some opening discussion toward tying scientific knowledge to belief.  
\end{abstract}


\maketitle
  
\section{Introduction}\label{intro}

Locating, structuring, thereafter simulating stylized facts on the diffusion of knowledge becomes increasingly difficult  (see e.g.  \cite{NKVMRAch3book}) due to the huge accumulation of big data. Therefore it is quite needed to downsize the investigations  in order to pin point "microscopic phenomena" contributing to the formation of "macroscopic  features". This process of looking at the nonlinear dynamics of interacting  intelligent populations \cite{nonl4}, in socio-physics,  is equivalent to the observation of vortices or solitons,  in fluid mechanics \cite{syst4,vx5}, before attempting to observe and to describe turbulence \cite{compl5}.

It is  nowadays commonly accepted that analyzing and modeling real-world phenomena  can be made on  
complex networks  \cite{costa0711.3199v2}.
In recent times, interesting observations on the diffusion of knowledge in structuring time dependent networks have followed such a path \cite{scholtes2013slow,scholtes2014causality} . In that respect, it was shown how  time ordering interactions, thus causality, affect  the interpretation of dynamical processes: in particular, by comparing  contrasting features on  moderate size time aggregated networks and on  their sub-structured time dependent counterparts. Authors \cite{scholtes2013slow,scholtes2014causality} also  showed   that some community  detection can be made by means of spectral clustering.

Here, a sort of inverse approach is presented. Considering a well defined set of interactions on a network, it  will  be observed that a difference in the diffusion of knowledge occurs depending on the sub-network size and structure.  Numerical results  are presented  from the comparison of several (small) networks, either when the time ordering of nodes is taken into account or when the network is seen after some time aggregation.


In order to do so,   several networks,  approximately of  the same size, but containing different types of sites and links have been  studied.  These networks look like
small world  networks.  
Moreover, it is imposed that the nodes belong to two communities made of stubborn agents, in order to keep a systematic topological structure, i.e. the diffusion  of knowledge is supposed to exist, but without a modification of the state of the recipient, - as  when insults are exchanged between agents.  Such communities having markedly different opinions have been previously studied in general frameworks  \cite{EJSP21.91.49 galamMosco,PHA381.07.366galam,PRE87.13.042807galammartins}. Such communities are, for example, the  Neocreationist and Intelligent Design  Proponents (IDP), on one hand, and the Darwinian  Evolution Defenders (DED),  on the other hand \cite{ARA26scott,greenwalt2003establishing}. Previous reports on these communities studied along the lines of  opinion formation,   as well as of behavior choice and agent  reactions   \cite{sznajd2000opinion,opinionformation}, within a  socio-physics  context pertaining to the  diffusion of ideas have been presented and are very briefly recalled in Sect. \ref{previousworks}.

The case of knowledge diffusion within such communities is studied  from time ordered selected quotations of agents, whence  after  building     networks, each   mimicked by its  adjacency matrix, with ranks and rows ordered to define  inter- and intra-community links. These networks are  intrinsically directed and not necessarily reciprocal.   Thus, the adjacency matrices have complex eigenvalues, and the eigenvectors present complex components  \cite{GRMAEPJB86}.  The content of the citations is not studied, but perusal of these indicate that they are more "negative arguments" than "positive ones". However, the diffusion of "knowledge" exists, but without a modification of the state of the recipient, e.g. like when insults are exchanged in many social worlds. There is hardly a search for consensus in such "controversies", indeed.

In Sect. \ref{77x77},  the large  77x77 matrix, i.e. a 77 network, is presented. 
In Sect. \ref{32x32}, it is explained that several sub-networks can be extracted for further study: they correspond to small world  networks which  contain  either an $odd$ or an $even$ number of nodes, in order to pin point the  relevance of complex eigenvalues of the pertinent matrix,  due to triads of agents. In so doing, it might be possible   to observe some possible symmetry (or "transitivity") effects, if any.
 Thereafter, in Sect. \ref{Aggrmatrixconstruct},  the time aggregated (thus, undirected) network counterpart, - which has an adjacency matrix which  is symmetric, whence has necessarily only real eigenvalues,   is constructed and analyzed.

A quantification of   the slow-down or speed-up  effects of information diffusion in  such temporal networks, with non-Markovian contact sequences, can be made by comparing the real time dependent (directed) network to its counterpart, the time aggregated (undirected) network, - which has real eigenvalues; see Sect. \ref{results} and Sect. \ref{slowdown}.

  \section{Perspective on specific stubborn agents} \label{previousworks} 

 With the aim of capturing the dynamical aspects of the interaction between   Neocreationist and Intelligent Design  Proponents (IDP) and the Darwinian  Evolution Defenders (DED),   agents of the IDP and DED groups,    the {\it degree of activity} of each group and the corresponding {\it degree of impact} on the   community can be monitored  \cite{garcia,tierney2008intelligent}.  From a mere opinion formation point of view, it could be shown that  if DED would have simply outlined scientific  data, i.e., not stating Darwin theory is "proved", but instead noting that it is the best frame to date, they would have  lost the debate against the
IDP   \cite{pha389.10.3619 galam,PhA390.11.3036galam}. 
 
In order to gain insight on the degree of interrelations due to the  activity of  such antagonistic  social groups \cite{palevitz2000falling,pizzo2006communication},  a  directed network  of citations can be constructed,   by applying   the  procedure  found in \cite{garcia} and recalled in Sect. \ref{construction}. 

\subsection{Network construction}\label{construction}

In order to build the network,the main agents  of the Intelligent
Design (ID) movement were first selected. From a paper by
R. T. Pennock \cite{Pennock},  criticizing ID,   the founders of the ID
movement are first identified.   Next, the ID web pages and their corresponding links were examined, starting from the URL of the Discovery InstituteÕs Center for
Science and Culture (CSC)  \cite{CSC}. Thereafter,
the connexions of  this predefined ID community  with the defenders of the other community, the
DarwinÕs evolution theory defenders are selected. This is helped by considering the increasing impact of the ID movement has impelled. the latter has by reaction activated  social and scientific organizations around the
world. Among the most important ones, the non-profit organization
National Center for Science Education (NCSE)  \cite{NCSE} plays a
relevant role in coordinating the activity of people defending the
teaching of evolutionary biology in the USA. 

A citation
network  has  been constructed as follows, in brief searching for nodes (agents of any community citing their own community or the opponents):
\begin{itemize}
\item  starting from a list containing the name of some of
the IDP  W. Dembski, M. Behe and S. Meyer, and using  Google  Scholar  Internet search tool, 
their  main publications were selected
\item next  another list   was created with the different authors citing
the agents of the previous list, while as objectively as possible recording their general
positions upon either one of the two sides of the debate
\item   a  node number was arbitrarily given to each agent
\item the node  or agent was  endowed  with an attribute according to the
apparent community position 
\item for each pair of  agents  a directed link was drawn
 if, according to the outcome of the Google  Scholar search process, there is a citation
\end{itemize}

N.B.  the data was downloaded and examined between Oct. 01 and Nov 15, 2007.  

\subsection{Network characteristics}
The network is composed of two subgraphs, one  with  37 and the other with  40 agents, corresponding to IDP and DED communities, respectively.  There are 170 and 128 links in IDP and  DED intra-communities, respectively, and 217  inter-community   links. Notice that no weight  is given to any  link.
 One can distinguish between {\it  directed links} (DL) and  {\it undirected links} (UL): such  a latter link connects a pair of nodes in both directions (A cites B and B cites A); by extension,  a  {\it Directed Triangle} (DT) is the shortest cycle of a graph formed by ONLY directed links (DL) 
(A cites B cites C cites A, but B does not cite A, etc.). However, the set-up of  such adjacency matrices is such that it is impossible to  report whether A cited B, before or after C cited B, for example.  Moreover, it is also obvious that a  DT is a chronologically  absurd feature, - except if there are  different citations, but this is not recorded in the present procedure. Nevertheless, the adjacency matrices are usually non-symmetric. Thus, systems of unspecified  (i.e., directed or undirected) links are  at first only those to be considered. One should remind the reader here that  the sum of an adjacency matrix and of its transpose, leads to a symmetric  matrix with different  weights $w$  for directed ($w=1$) and  for undirected  ($w=2$) links

 Nevertheless, individuals leading the transfer of opinion between  IDP and DED communities can be  identified by analyzing the number of directed triangles and of undirected links of the citation network.  
It was found  \cite{garcia} that the three main nodes in the ID community make up for 56$\%$ of the IDP triangles and 41 $\%$ of the inhomogeneous ones, while  5 nodes in the DE community make up for 51$\%$ of the inhomogeneous triangles.
Thus it can be safely assumed that a few so called opinion leaders can well describe the activity of   the whole group to which they belong.

Whence, it should be obvious to the reader that in such small networks, it is hard to get a convincing power law of the degree distributions. However,  it  seems easily induced that the preferential attachment mode is pertinent for each community, with different  "scholarly constraints", as it will be deduced in Sect. \ref{results} and argued upon in Sect. \ref{conclusion}.

  \begin{table}
\caption{Number of nodes, links, and triads,  with unspecified  (directed or undirected) edges between indicated  types of nodes, in the  various networks represented by the  various adjacency matrices of indicated size 
}
\smallskip
\begin{footnotesize}
\begin{center}
\begin{tabular}{|c| c| c ccc cccc|}
  \hline
  Matrices (Networks)	& $M_{77}$  &$C_{12}$ & $D_{12}$&$M_{24}$& $C_{14}$& $D_{15}$& $M_{29}$& $F_{24}$&$F_{29}$ \\ \hline
  $N. nodes$ & 77& 12 &12 &24  &14 &15 &29 &24   &29 \\\hline
  $N. links$ & 307& 27 &37 &125  &35 &46 &152&46 &54 \\\hline
  \hline 
 &\multicolumn{9}{|c|}{Number of triads } \\ \hline
 Triad configuration	& $M_{77}$  &$C_{12}$ & $D_{12}$&$M_{24}$& $C_{14}$& $D_{15}$& $M_{29}$
 && \\ \hline
   IDP-IDP-IDP 	& 21& 12 &-& 12&18&-&18  &&
\\\hline   IDP-IDP-DED 	& 105&-&-&88&-&-&99 &&
\\\hline   IDP-DED-DED	& 171&-&-&129&-&-&165  &&
\\\hline   DED-DED-DED 	& 51&-&22&22&-&28&28&&
\\\hline
\end{tabular}
\label{tab01} 
\end{center}
\end{footnotesize}
\end{table}


  It is worth calculating the   number of  
   triangles associated to each group, i.e. depending on the type of nodes on the triangle edges.  The results for the different possible types of triangles are given in Table I. We emphasize that triangles containing elements of different  communities are the most abundant ones.  Conversely, among the 348 triangles sustaining the network, 72 (thus 0.21 $\%$) are homogeneous, relating only nodes of the same community.  Thus, it is obvious that the  interactions induce some non trivial dynamics    \cite{physaGR,GRMAEPJB86}, leading to real and complex eigenvalues.

 \begin{table} \begin{center}
\caption{Characteristics of the main eigenvalues (EVs) for   the  various  adjacency matrices pertinent  to  the investigated networks. the cases in which the modulus of an EV (but not the largest one)  is larger than the modulus of the second largest real EV  
is emphasized   
with an !!}
\smallskip
\begin{footnotesize}
\begin{tabular}{|c| c|  c|  c| c| c|  c| c| c|c| }
\hline 
&\multicolumn{9}{|c|}{Matrix } \\ \hline
 	& $M_{77}$  &$C_{12}$& $D_{12}$&$M_{24}$& $C_{14}$& $D_{15}$& $M_{29}$& $F_{24}$&$F_{29}$ 
\\\hline  N. of  Re EVs 	& 14&2&2&4&2&2&4&5&2
\\\hline N. of   c.c.  EVs	&12&1&3&7&1&3&8&0& 2   
\\\hline   Largest EV ($\lambda_1$)	&7.994&2.588&3.591&6.777&2.588&3.591&6.845&3.661&3.764
\\\hline  $\lambda_2$ if Re EV 	& 2.481&-1.742&1.432&1.553&-1.742&1.432&1.596&0.773&0.662
\\\hline    $| \lambda_2| $ & 2.481&1.742&1.482 !!&1.827 !!&1.742&1.482 !!&1.760 !!&3.661 !!&3.764 !!
\\\hline due to&& &-1.462 & -1.577 &&-1.462 &   -1.583 &-3.661  &-3.764
\\ && & $\pm$ $\;i$ 0.240 & $\pm$ $\;i$ 0.923 && $\pm$ $\;i$ 0.240& $\pm$ $\;i$ 0.768 &&\\ \hline
$| \lambda_2| /\lambda_1$&0.3104 &0.6731 &0.412&0.2696 &0.6731 &0.4127 &0.2571 & 1.& 1.   \\\hline
\hline  	& $\widehat{M_{77}}$  &$\widehat{C_{12}}$ &$ \widehat{D_{12}}$&$\widehat{M_{24}}$&$\widehat{C_{14}}$&$\widehat{D_{15}}$&$\widehat{M_{29}}$&$\widehat{F_{24}}$&$\widehat{F_{29}}$ 
\\\hline   N. of  Re EVs  $\neq0$	& 61&11&11&24&12&15&29&22&26
\\\hline   N. of    EVs = 0	& 16 &1 &1&0&2&0&0&2&3
\\\hline   Largest EV 	& 8.226&3.089&3.9285&7.2485&3.288&4.072&7.450&4.0595&4.1905
\\\hline   $\tilde{\lambda}_2$ if  Re EV 	&  2.771&1.1615&1.5530&1.9845&1.4895&1.8745&2.305&1.3785&1.4745
\\\hline     $|\tilde{\lambda}_2| $ 	&3.824 !!&2.050&1.845 !!&2.5635&2.078 !!&1.920 !!&2.748 !!&4.0595&4.1905 !!
\\\hline \hline
 $|\tilde{\lambda}_2| $/ $|\tilde{\lambda}_1| $ & 0.4648&0.6636 &0.4696&0.3537 &0.6320 &0.4715 &0.3689 &1. & 1. \\\hline 
 \hline   $S^*$(Eq.\ref{S})	&1.350&1.0359  &  0.8523& 0.7930 &1.1592  & 0.8495 &0.7343  & - & -
\\\hline
\end{tabular}
\label{SummaryEVEvof8matrices}
\end{footnotesize} \end{center}
\end{table}

    \subsection{A 77x77 real asymmetric matrix } \label{77x77}
  
 The adjacency matrices can be summarized as

\begin{equation} \label{M0}
M_0\equiv\left( \begin{tabular}{ll} $ C_0$  &   $\;A$   \\  $ B $ & $D_0$ 
\end{tabular} \right)\; \equiv \left( \begin{tabular}{ll} $ C_0$  &   $\;0$   \\  $ 0 $ & $D_0$ 
\end{tabular} \right)+  \left( \begin{tabular}{ll} 0 & $A$  \\  $B$ & $\;0$\end{tabular} \right).
\end{equation} in which a matrix element $m_{ij}$ takes the value 1 or 0 depending on
whether or not  a citation of $i$ by $j$  has taken place, 
as recorded and explained in ref.  \cite{garcia,physaGR}. The matrices   $C_0$  (37x37) and $D_0$ (40x40)  indicate whether  agents of community $i$  have been quoted by others of the same community $i$. Self-citations are disregarded, $m_{ii}=0$, i.e. all diagonal terms in $M_0$, $C_0$, and $D_0$ are $0$;  see   \cite{physaGR} for the list of all finite matrix elements. 
 In contrast,  $F_0$, i.e.
\begin{equation} \label{F}
F_0=\left( \begin{tabular}{ll} 0 & $A$  \\  $B$ & $\;0$
\end{tabular} \right).
\end{equation}  emphasizes links $between$  different communities, i.e.    agents of community $j$  quoting those of community $i(\neq j)$; $i \leftarrow j$.
 $A$ and $B$ are obviously $rectangular$ matrices describing inter-community  links. $A$ and $B$ are (40x37) and (37x40) matrices respectively. All  $C_0$, $A$, $B$ and $D_0$  matrices are given in the Appendix; from such matrices, the network, not shown for space saving, can be easily reconstructed through any good classical graph software. 
 

Moreover,   since each square matrix  $M_0$, $C_0$, $D_0$, $F_0$ has non-negative elements,  the Perron-Frobenius theorem  states that there exists  at least one non-negative eigenvalue greater or equal  {\it in absolute value} than all other eigenvalues;  its 
  corresponding eigenvector has non-negative components  \cite{Meyer 2000,Gantmacher 2000}. 


The Perron-Frobenius theorem, applied in its version for non-negative matrices only, indicates that there may exist eigenvalues of the same absolute value as the maximal one. 

 The EVs of the above 4 matrices,  $M_0$, $C_0$,  $D_0$, and $F_0$,  have been computed. The EVs of interest are given in  
 Table \ref{SummaryEVEvof8matrices}. In the framework of this paper only these relevant EVs are indicated. 
 
 \subsection{ Reduced size "core networks"} \label{32x32}
 
To proceed on knowledge diffusion on a network, it should be recognized that some knowledge is lost when the agent is not well connected, i.e. has a node degree = 1, either being at the end  of a "dangling bond", - like a sink or a source in fluid mechanics.  The same can be thought for a node degree =  2.  The role of these  agents  is likely marginal in contrast to those sinks and sources which are hubs of the network. In fact for the diffusion of knowledge, a triad graph seems the most basic graph to consider. Thus, for the present  study, a few nodes of the whole network can be eliminated from the start.
 
The following procedure has been applied: in order to emphasize the role of  inter-community connecting agents,   all agents in the $A$ and $B$ rectangular matrices  are kept when  $a_{i,j}=1$  and $b_{i,j}=1$, whence reducing the entire community network to its relevant core according to inter-community links. Next,  the most important agents in the $C$ and $D$ matrices are conserved as nodes relevant for intra-community knowledge transfer.  
Remembering that  an odd or an even number of nodes might lead to different sets of  (real or complex) eigenvalues, it is interesting to compare two related networks  different by only one  node unit.  It results that   the following cases  are be considered:
\begin{itemize}
\item a network made of  12 ISP
\item a network made of  12 DED
\item a network made of   24 nodes : 12 DED and 12 IDP
 \item  a network made of  14 IDP
\item a network made of  15 DED
\item a network made of  29 nodes : 14 IDP and 15  DED
\end{itemize}
 
Each network is represented by its adjacency matrix: $C_{12}$, $D_{12}$, $M_{24}$, $C_{14}$, $D_{15}$,  and $M_{29}$. Moreover, in order to investigate further any effect in the inter community "knowledge sharing",   the corresponding $F$ matrices, i.e. $F_{24}$ and $F_{29}$ have to be specifically used. Recall that none of these 8 matrices are symmetric.

 \subsection{Aggregation matrix (or network)  construction} \label{Aggrmatrixconstruct}
 
 The above networks, reproducing citations, contain some directed links (and UL as well). For each considered  ("time-dependent")  network,   an "aggregated  network" can be defined though  a (new)  adjacency matrix, e.g. $M_{i,j} \rightarrow $ 
 $\widehat{M_{i,j}}$, etc., - for the 8 cases outlined here above.  These new 8  $\widehat{M_{i,j}}$ matrices are necessarily symmetric.
 
 \section{Results}\label{results}
 
 For each (16) matrix,  the eigenvalues (and eigenvectors) have been    calculated. The main characteristic results relevant to the present investigation 	are given in Table \ref{SummaryEVEvof8matrices}. To distinguish between
  the number of  real EVs  $\neq0$	or   0  is not  the presently  relevant subject. However, it is at once pointed out that the largest EV is of course real and positive in each case, but  the magnitude of  the most negative EV might be larger than the
"Next to Largest" Re  (positive) EV, {\it a fortiori} if the latter is negative. Thus, in   Table \ref{SummaryEVEvof8matrices}, and in view of preparing the following sections and discussion, both the strictly  "Next to Largest" Re  (positive) EV is given but also the EV having the
"Next to Largest Modulus". It can be seen from this Table that the networks containing DED are those for which the distinction on the notion of "next to largest" EV is relevant.
 
 \section{Slow-down or Speed-up knowledge diffusion}\label{slowdown}
 
 It has been shown  that changes of diffusion dynamics in temporal networks  as compared to their static counterparts are due to the change of connectedness or conductance of the corresponding second-order aggregate network. These changes influence the  process of knowledge diffusion through a slow-down or speed-up factor which can be computed based on the second-order aggregate networks corresponding to a particular non-Markovian temporal network and its Markovian counterpart. This basically consists in comparing two corresponding adjacency matrix features. In the present case, where the usual temporal aspects is masked but replaced by a sequential one (of quotations), the matter consists in comparing the original adjacency matrix and its symmetrized counterpart.

 

 It was interestingly shown  that  the convergence time of random walks
is related to the second largest eigenvalue of the transition matrix $T $. For a primitive stochastic
matrix with (not necessarily real) eigenvalues $ 1 = \lambda_1 >  |\lambda_2| >   |\lambda_3|   \ge \dots  \ge   |\lambda_n|   $,  it was 
shown that the number of steps $k$ after which the total variation distance $ \Delta(\pi_k;\pi)$ between the
visitation probabilities  $\pi_k$ and the stationary distribution $\pi$ of a random walk falls below $\epsilon$  is
proportional to  $1/ ln(|\lambda_2|)$.
  For a matrix $ T^{(2)}$ capturing the statistics of two-paths in an empirical temporal
network 
 and a matrix $\tilde{T}^{(2)}$  representing the ÒMarkovianÓ model derived from the
 symmetrized network,  
an analytical prediction for the change of convergence speed   $ S^*$, due to non-Markovian properties can be derived as
 \begin{equation}\label{S} S^*(T^{(2)}) := ln(|\tilde  \lambda_2|)/ ln(| \lambda_2|) 
 \end{equation}
 where $ \lambda_2$ and $\tilde \lambda_2$ denote the second largest eigenvalue of $T^{(2)}$ and $\tilde{T}^{(2)}$  respectively.   Thus, a diffusion slow-down exists if
$ S^*(T^{(2)}) \ge1$.
A diffusion  speed-up exists if 
 $S^*(T^{(2)})  \le 1$.    To calculate   $S^*(T^{(2)}) $, in the present cases, observe that  Eq. (\ref{S}) must be adapted to take into account the ("normalizing") $\lambda_1$  value; see {\it ad hoc} line in    Table \ref{SummaryEVEvof8matrices}.
 
 
 
 In the present network cases, a temporal network  adjacency matrix can have its "second  largest eigenvalue", i.e. 
  to be considered as the "next to  largest" eigenvalue,  
 either real  (positive or negative, in fact) or  be a c.c. eigenvalue with a  large modulus.

   The relevant results are given in   Table \ref{SummaryEVEvof8matrices} last line. For a global view of  the data, one can rank the $S^*(T^{(2)}) $ values in decreasing order: this corresponds to rank the networks as follows: $M_{77}$,  $C_{14}$,  $C_{12}$,  which have a slow-down feature, while  $D_{12}$,  $D_{15}$,  $M_{24}$,  and $M_{29}$ possess a speed-up feature.  
   
   It is deduced that  \begin{itemize} 
   \item (i) the diffusion of knowledge is more difficult on the large (complete) network, but this could have been expected;
   \item (ii)  the same type of  hierarchy constraint  on the network size is found either for the slowing-down or speeding up processes;
   \item (iii) however, the IDP and DED sub-networks are markedly different: the diffusion of knowledge is slower for IDP, but faster for DED; this  ({\it a priori}  unexpected finding) might nevertheless be "rationalized", if  one attempts to introduce some "level of scientific quality" in the behavior of the various agents. This perspective offers some opening discussion toward tying psychology, intellect, scientific knowledge to belief. However, one cannot completely neglect the fact  that the DED might have more use in publishing thoughts than IDP, who might be less prone to practically publish, whence be quoted; 
   \item (iv) another interesting point pertains to the relative influence of the agents on the  (reduced, but pertinent) networks:   the diffusion of knowledge is  markedly in favor of the DED, since the  $M_{24}$ and $M_{29}$ corresponding speeds are obviously  on the  up  side.
   \end{itemize}

 
\section{Conclusion}\label{conclusion}

As a conclusion, let a brief summary be given tying the "questions" to the "answers". In the main text,  it has been studied whether the diffusion of knowledge can be measured in and outside distinct communities, necessarily made of stubborn agents on small world-like networks. This  speed of knowledge diffusion is obtained from the eigenvalues of the corresponding adjacency matrices  for the whole set  of agents and for their sub-communities.  In particular, it has been found that the  Neocreationist and Intelligent Design  Proponents (IDP), on one hand, and the Darwinian  Evolution Defenders (DED),  on the other hand behave quite differently in processing the knowledge.  A  quantification of   the slow-down or speed-up  effects of information diffusion in  such temporal networks, with non-Markovian contact sequences, has been made.  It is observed that the  diffusion of knowledge is slower in IDP and  faster in DED communities. It is  argued that the finding can be "rationalized", if   some "scientific quality"  and "publication habit" are attributed to the agents, as common sense would suggest.   This  finding offers some opening discussion toward tying scientific knowledge to belief, and subsequent diffusion of both in small worlds.  

Moreover, a brief observation has been made on the community size effect, and its substructure.  It is observed that  the diffusion of knowledge is more difficult on  large  networks. It is also observed that the number of triads with heterogeneous agents seems a relevant "parameter". In the present cases, a speed-up process effect is markedly greater when two DED agents are involved,  whence again likely pointing to some behavior origin in the more usual scientific arguing methods  prone to such a community. Since it has been found  in   \cite{GRMAEPJB86} that the origin of complex eigenvalues is related to the structure of triads, further work on the relationship between the  (density of) different types of triads and the speed of  knowledge diffusion should be interesting.

 \begin{flushleft}
{\large \bf Acknowledgment}
\end{flushleft}
 This paper is part of scientific activities in COST Action TD1210   'Analyzing the dynamics of information and knowledge landscapes'.

 \section*{Appendix   IDP-DED matrix}\label{sec:App_matrix}

In this Appendix,  the adjacency matrices of interest, $C_0$, $A$, $B$, $D_0$, are recalled.

\begin{table}\label{table:C0} \begin{tabular}{|l|l|l|l|l|} \hline $C_0$
  &
   1
 $\cdots$ &$\cdots$&$\cdots$&$\cdots$ \hspace{37pt} ... 37
 \\ \hline
 1& 0  1  1  1   0\hspace{1pt}   0\hspace{1pt}   0\hspace{1pt}   0\hspace{1pt}   0\hspace{1pt} & 1   0\hspace{1pt}  1  1  1  1  1   0\hspace{1pt}   0\hspace{1pt}   0\hspace{1pt} &  0\hspace{1pt}  1   0\hspace{1pt}  1   0\hspace{1pt}   0\hspace{1pt}   0\hspace{1pt}   0\hspace{1pt}   0\hspace{1pt}  1 &  0\hspace{1pt}  1   0\hspace{1pt}   0\hspace{1pt}   0\hspace{1pt}   0\hspace{1pt}  1   0\hspace{1pt} \\  
 2& 1  0  1   0\hspace{1pt}   0\hspace{1pt}   0\hspace{1pt}   0\hspace{1pt}   0\hspace{1pt}   0\hspace{1pt} &  0\hspace{1pt}   0\hspace{1pt}  1  1  1  1  1   0\hspace{1pt}   0\hspace{1pt}   0\hspace{1pt} &  0\hspace{1pt}  1   0\hspace{1pt}  1   0\hspace{1pt}   0\hspace{1pt}   0\hspace{1pt}   0\hspace{1pt}   0\hspace{1pt}  1 &  0\hspace{1pt}  1  1   0\hspace{1pt}   0\hspace{1pt}   0\hspace{1pt}  1   0\hspace{1pt} \\  
 3& 1   0\hspace{1pt}   0\hspace{1pt}   0\hspace{1pt}   0\hspace{1pt}   0\hspace{1pt}   0\hspace{1pt}  1   0\hspace{1pt} &  0\hspace{1pt}   0\hspace{1pt}   0\hspace{1pt}   0\hspace{1pt}   0\hspace{1pt}   0\hspace{1pt}   0\hspace{1pt}   0\hspace{1pt}  1   0\hspace{1pt} & 1  1   0\hspace{1pt}   0\hspace{1pt}   0\hspace{1pt}   0\hspace{1pt}   0\hspace{1pt}   0\hspace{1pt}  1   0\hspace{1pt} &  0\hspace{1pt}   0\hspace{1pt}   0\hspace{1pt}   0\hspace{1pt}   0\hspace{1pt}   0\hspace{1pt}   0\hspace{1pt}   0\hspace{1pt} \\  
 4& 1   0\hspace{1pt}   0\hspace{1pt}   0\hspace{1pt}   0\hspace{1pt}   0\hspace{1pt}   0\hspace{1pt}   0\hspace{1pt}   0\hspace{1pt} &  0\hspace{1pt}   0\hspace{1pt}   0\hspace{1pt}   0\hspace{1pt}   0\hspace{1pt}   0\hspace{1pt}   0\hspace{1pt}   0\hspace{1pt}   0\hspace{1pt}  1 & 1  1   0\hspace{1pt}   0\hspace{1pt}   0\hspace{1pt}   0\hspace{1pt}   0\hspace{1pt}   0\hspace{1pt}   0\hspace{1pt}   0\hspace{1pt} &  0\hspace{1pt}   0\hspace{1pt}   0\hspace{1pt}   0\hspace{1pt}   0\hspace{1pt}   0\hspace{1pt}   0\hspace{1pt}   0\hspace{1pt} \\  
 5& 1   0\hspace{1pt}  1  1   0\hspace{1pt}  1   0\hspace{1pt}  1  1 &  0\hspace{1pt}   0\hspace{1pt}   0\hspace{1pt}   0\hspace{1pt}   0\hspace{1pt}   0\hspace{1pt}   0\hspace{1pt}   0\hspace{1pt}   0\hspace{1pt}   0\hspace{1pt} &  0\hspace{1pt}   0\hspace{1pt}  1   0\hspace{1pt}   0\hspace{1pt}   0\hspace{1pt}   0\hspace{1pt}   0\hspace{1pt}   0\hspace{1pt}   0\hspace{1pt} &  0\hspace{1pt}   0\hspace{1pt}   0\hspace{1pt}   0\hspace{1pt}   0\hspace{1pt}   0\hspace{1pt}   0\hspace{1pt}   0\hspace{1pt} \\  
 6& 1   0\hspace{1pt}  1   0\hspace{1pt}   0\hspace{1pt}  0   0\hspace{1pt}  1   0\hspace{1pt} &  0\hspace{1pt}   0\hspace{1pt}   0\hspace{1pt}   0\hspace{1pt}   0\hspace{1pt}   0\hspace{1pt}   0\hspace{1pt}   0\hspace{1pt}   0\hspace{1pt}   0\hspace{1pt} &  0\hspace{1pt}   0\hspace{1pt}   0\hspace{1pt}  1   0\hspace{1pt}   0\hspace{1pt}   0\hspace{1pt}   0\hspace{1pt}   0\hspace{1pt}   0\hspace{1pt} &  0\hspace{1pt}   0\hspace{1pt}   0\hspace{1pt}   0\hspace{1pt}   0\hspace{1pt}   0\hspace{1pt}   0\hspace{1pt}   0\hspace{1pt} \\  
 7& 1   0\hspace{1pt}   0\hspace{1pt}   0\hspace{1pt}   0\hspace{1pt}   0\hspace{1pt}   0\hspace{1pt}   0\hspace{1pt}   0\hspace{1pt} &  0\hspace{1pt}   0\hspace{1pt}   0\hspace{1pt}   0\hspace{1pt}   0\hspace{1pt}   0\hspace{1pt}   0\hspace{1pt}  1   0\hspace{1pt}   0\hspace{1pt} &  0\hspace{1pt}   0\hspace{1pt}   0\hspace{1pt}   0\hspace{1pt}  1  1  1   0\hspace{1pt}   0\hspace{1pt}   0\hspace{1pt} &  0\hspace{1pt}   0\hspace{1pt}   0\hspace{1pt}   0\hspace{1pt}   0\hspace{1pt}   0\hspace{1pt}   0\hspace{1pt}   0\hspace{1pt} \\  
 8&  0\hspace{1pt}  1   0\hspace{1pt}  1   0\hspace{1pt}   0\hspace{1pt}   0\hspace{1pt}  0   1 &  0\hspace{1pt}   0\hspace{1pt}   0\hspace{1pt}   0\hspace{1pt}   0\hspace{1pt}   0\hspace{1pt}   0\hspace{1pt}   0\hspace{1pt}   0\hspace{1pt}   0\hspace{1pt} &  0\hspace{1pt}   0\hspace{1pt}   0\hspace{1pt}   0\hspace{1pt}   0\hspace{1pt}   0\hspace{1pt}   0\hspace{1pt}  1  1  1 & 1  1   0\hspace{1pt}   0\hspace{1pt}   0\hspace{1pt}   0\hspace{1pt}   0\hspace{1pt}   0\hspace{1pt} \\  
 9&  0\hspace{1pt}   0\hspace{1pt}   0\hspace{1pt}   0\hspace{1pt}   0\hspace{1pt}   0\hspace{1pt}   0\hspace{1pt}   0\hspace{1pt}   0\hspace{1pt} &  0\hspace{1pt}   0\hspace{1pt}   0\hspace{1pt}   0\hspace{1pt}   0\hspace{1pt}   0\hspace{1pt}   0\hspace{1pt}   0\hspace{1pt}   0\hspace{1pt}   0\hspace{1pt} &  0\hspace{1pt}   0\hspace{1pt}   0\hspace{1pt}   0\hspace{1pt}   0\hspace{1pt}   0\hspace{1pt}   0\hspace{1pt}   0\hspace{1pt}   0\hspace{1pt}   0\hspace{1pt} &  0\hspace{1pt}   0\hspace{1pt}  1  1   0\hspace{1pt}   0\hspace{1pt}   0\hspace{1pt}   0\hspace{1pt} \\  
10&  0\hspace{1pt}
 0\hspace{1pt}
 0\hspace{1pt}
 0\hspace{1pt}
 0\hspace{1pt}
 0\hspace{1pt}
 0\hspace{1pt}
 0\hspace{1pt}
 0\hspace{1pt} &
 0\hspace{1pt}
 0\hspace{1pt}
 0\hspace{1pt}
 0\hspace{1pt}
 0\hspace{1pt}
 0\hspace{1pt}
 0\hspace{1pt}
 0\hspace{1pt}
 0\hspace{1pt}
 0\hspace{1pt} &
 0\hspace{1pt}
 0\hspace{1pt}
 0\hspace{1pt}
 0\hspace{1pt}
 0\hspace{1pt}
 0\hspace{1pt}
 0\hspace{1pt}
 0\hspace{1pt}  1
 0\hspace{1pt} &
 0\hspace{1pt}
 0\hspace{1pt}  1
 0\hspace{1pt}  1  1
 0\hspace{1pt}
 0\hspace{1pt} \\   11&
 0\hspace{1pt}
 0\hspace{1pt}
 0\hspace{1pt}
 0\hspace{1pt}
 0\hspace{1pt}
 0\hspace{1pt}
 0\hspace{1pt}
 0\hspace{1pt}
 0\hspace{1pt} &
 0\hspace{1pt}
 0\hspace{1pt}
 0\hspace{1pt}
 0\hspace{1pt}
 0\hspace{1pt}
 0\hspace{1pt}
 0\hspace{1pt}  1
 0\hspace{1pt}
 0\hspace{1pt} &
 0\hspace{1pt}
 0\hspace{1pt}
 0\hspace{1pt}
 0\hspace{1pt}
 0\hspace{1pt}
 0\hspace{1pt}
 0\hspace{1pt}
 0\hspace{1pt}
 0\hspace{1pt}
 0\hspace{1pt} &
 0\hspace{1pt}  1
 0\hspace{1pt}
 0\hspace{1pt}
 0\hspace{1pt}
 0\hspace{1pt}  1  1 \\   12&
 0\hspace{1pt}
 0\hspace{1pt}
 0\hspace{1pt}
 0\hspace{1pt}
 0\hspace{1pt}
 0\hspace{1pt}
 0\hspace{1pt}
 0\hspace{1pt}
 0\hspace{1pt} &
 0\hspace{1pt}
 0\hspace{1pt}
 0\hspace{1pt}
 0\hspace{1pt}
 0\hspace{1pt}
 0\hspace{1pt}
 0\hspace{1pt}
 0\hspace{1pt}
 0\hspace{1pt}
 0\hspace{1pt} &
 0\hspace{1pt}
 0\hspace{1pt}
 0\hspace{1pt}
 0\hspace{1pt}
 0\hspace{1pt}
 0\hspace{1pt}
 0\hspace{1pt}
 0\hspace{1pt}
 0\hspace{1pt}
 0\hspace{1pt} &
 0\hspace{1pt}
 0\hspace{1pt}
 0\hspace{1pt}
 0\hspace{1pt}
 0\hspace{1pt}
 0\hspace{1pt}
 0\hspace{1pt}
 0\hspace{1pt} \\   13&
 0\hspace{1pt}
 0\hspace{1pt}  1
 0\hspace{1pt}
 0\hspace{1pt}
 0\hspace{1pt}
 0\hspace{1pt}
 0\hspace{1pt}
 0\hspace{1pt} &
 0\hspace{1pt}
 0\hspace{1pt}
 0\hspace{1pt}  0
 0\hspace{1pt}
 0\hspace{1pt}
 0\hspace{1pt}
 0\hspace{1pt}
 0\hspace{1pt}
 0\hspace{1pt} &
 0\hspace{1pt}
 0\hspace{1pt}
 0\hspace{1pt}
 0\hspace{1pt}
 0\hspace{1pt}
 0\hspace{1pt}
 0\hspace{1pt}
 0\hspace{1pt}
 0\hspace{1pt}
 0\hspace{1pt} &
 0\hspace{1pt}
 0\hspace{1pt}
 0\hspace{1pt}
 0\hspace{1pt}
 0\hspace{1pt}
 0\hspace{1pt}
 0\hspace{1pt}
 0\hspace{1pt} \\   14&
 0\hspace{1pt}
 0\hspace{1pt}
 0\hspace{1pt}
 0\hspace{1pt}
 0\hspace{1pt}
 0\hspace{1pt}
 0\hspace{1pt}
 0\hspace{1pt}
 0\hspace{1pt} &
 0\hspace{1pt}
 0\hspace{1pt}
 0\hspace{1pt}
 0\hspace{1pt}
 0\hspace{1pt}
 0\hspace{1pt}
 0\hspace{1pt}
 0\hspace{1pt}
 0\hspace{1pt}
 0\hspace{1pt} &
 0\hspace{1pt}
 0\hspace{1pt}
 0\hspace{1pt}
 0\hspace{1pt}
 0\hspace{1pt}
 0\hspace{1pt}
 0\hspace{1pt}
 0\hspace{1pt}
 0\hspace{1pt}
 0\hspace{1pt} &
 0\hspace{1pt}
 0\hspace{1pt}
 0\hspace{1pt}
 0\hspace{1pt}
 0\hspace{1pt}
 0\hspace{1pt}
 0\hspace{1pt}
 0\hspace{1pt} \\   15& 1  1
 0\hspace{1pt}
 0\hspace{1pt}
 0\hspace{1pt}
 0\hspace{1pt}
 0\hspace{1pt}
 0\hspace{1pt}
 0\hspace{1pt} &
 0\hspace{1pt}
 0\hspace{1pt}
 0\hspace{1pt}
 0\hspace{1pt}
 0\hspace{1pt}  0
 0\hspace{1pt}
 0\hspace{1pt}
 0\hspace{1pt}
 0\hspace{1pt} &
 0\hspace{1pt}
 0\hspace{1pt}
 0\hspace{1pt}
 0\hspace{1pt}
 0\hspace{1pt}
 0\hspace{1pt}
 0\hspace{1pt}
 0\hspace{1pt}
 0\hspace{1pt}
 0\hspace{1pt} &
 0\hspace{1pt}
 0\hspace{1pt}
 0\hspace{1pt}
 0\hspace{1pt}
 0\hspace{1pt}
 0\hspace{1pt}
 0\hspace{1pt}
 0\hspace{1pt} \\   16& 1
 0\hspace{1pt}
 0\hspace{1pt}
 0\hspace{1pt}
 0\hspace{1pt}  1
 0\hspace{1pt}
 0\hspace{1pt}
 0\hspace{1pt} &
 0\hspace{1pt}
 0\hspace{1pt}
 0\hspace{1pt}
 0\hspace{1pt}
 0\hspace{1pt}
 0\hspace{1pt}
 0\hspace{1pt}
 0\hspace{1pt}
 0\hspace{1pt}
 0\hspace{1pt} &
 0\hspace{1pt}
 0\hspace{1pt}
 0\hspace{1pt}
 0\hspace{1pt}
 0\hspace{1pt}
 0\hspace{1pt}
 0\hspace{1pt}
 0\hspace{1pt}
 0\hspace{1pt}
 0\hspace{1pt} &
 0\hspace{1pt}
 0\hspace{1pt}
 0\hspace{1pt}
 0\hspace{1pt}
 0\hspace{1pt}
 0\hspace{1pt}
 0\hspace{1pt}
 0\hspace{1pt} \\   17&
 0\hspace{1pt}
 0\hspace{1pt}
 0\hspace{1pt}
 0\hspace{1pt}
 0\hspace{1pt}
 0\hspace{1pt}
 0\hspace{1pt}
 0\hspace{1pt}
 0\hspace{1pt} &
 0\hspace{1pt}
 0\hspace{1pt}
 0\hspace{1pt}
 0\hspace{1pt}
 0\hspace{1pt}
 0\hspace{1pt}
 0\hspace{1pt}
 0\hspace{1pt}
 0\hspace{1pt}
 0\hspace{1pt} &
 0\hspace{1pt}
 0\hspace{1pt}
 0\hspace{1pt}
 0\hspace{1pt}
 0\hspace{1pt}
 0\hspace{1pt}
 0\hspace{1pt}
 0\hspace{1pt}
 0\hspace{1pt}
 0\hspace{1pt} &
 0\hspace{1pt}
 0\hspace{1pt}
 0\hspace{1pt}
 0\hspace{1pt}
 0\hspace{1pt}
 0\hspace{1pt}
 0\hspace{1pt}
 0\hspace{1pt} \\   18&
 0\hspace{1pt}
 0\hspace{1pt}
 0\hspace{1pt}
 0\hspace{1pt}
 0\hspace{1pt}
 0\hspace{1pt}
 0\hspace{1pt}
 0\hspace{1pt}
 0\hspace{1pt} &
 0\hspace{1pt}
 0\hspace{1pt}
 0\hspace{1pt}
 0\hspace{1pt}
 0\hspace{1pt}
 0\hspace{1pt}
 0\hspace{1pt}
 0\hspace{1pt}
 0\hspace{1pt}
 0\hspace{1pt} &
 0\hspace{1pt}
 0\hspace{1pt}
 0\hspace{1pt}
 0\hspace{1pt}
 0\hspace{1pt}
 0\hspace{1pt}
 0\hspace{1pt}
 0\hspace{1pt}
 0\hspace{1pt}
 0\hspace{1pt} &
 0\hspace{1pt}
 0\hspace{1pt}
 0\hspace{1pt}
 0\hspace{1pt}
 0\hspace{1pt}
 0\hspace{1pt}
 0\hspace{1pt}
 0\hspace{1pt} \\   19&
 0\hspace{1pt}
 0\hspace{1pt}
 0\hspace{1pt}
 0\hspace{1pt}
 0\hspace{1pt}
 0\hspace{1pt}
 0\hspace{1pt}
 0\hspace{1pt}
 0\hspace{1pt} &
 0\hspace{1pt}
 0\hspace{1pt}
 0\hspace{1pt}
 0\hspace{1pt}
 0\hspace{1pt}
 0\hspace{1pt}
 0\hspace{1pt}
 0\hspace{1pt}
 0\hspace{1pt}
 0\hspace{1pt} &
 0\hspace{1pt}
 0\hspace{1pt}
 0\hspace{1pt}
 0\hspace{1pt}
 0\hspace{1pt}
 0\hspace{1pt}
 0\hspace{1pt}
 0\hspace{1pt}
 0\hspace{1pt}
 0\hspace{1pt} &
 0\hspace{1pt}
 0\hspace{1pt}
 0\hspace{1pt}
 0\hspace{1pt}
 0\hspace{1pt}
 0\hspace{1pt}
 0\hspace{1pt}
 0\hspace{1pt} \\   20&
 0\hspace{1pt}
 0\hspace{1pt}
 0\hspace{1pt}
 0\hspace{1pt}
 0\hspace{1pt}
 0\hspace{1pt}
 0\hspace{1pt}
 0\hspace{1pt}
 0\hspace{1pt} &
 0\hspace{1pt}
 0\hspace{1pt}
 0\hspace{1pt}
 0\hspace{1pt}
 0\hspace{1pt}
 0\hspace{1pt}
 0\hspace{1pt}
 0\hspace{1pt}  1
 0\hspace{1pt} &
 0\hspace{1pt}
 0\hspace{1pt}
 0\hspace{1pt}
 0\hspace{1pt}
 0\hspace{1pt}
 0\hspace{1pt}
 0\hspace{1pt}
 0\hspace{1pt}
 0\hspace{1pt}
 0\hspace{1pt} &
 0\hspace{1pt}
 0\hspace{1pt}
 0\hspace{1pt}
 0\hspace{1pt}
 0\hspace{1pt}
 0\hspace{1pt}
 0\hspace{1pt}
 0\hspace{1pt} \\   21&
 0\hspace{1pt}
 0\hspace{1pt}
 0\hspace{1pt}
 0\hspace{1pt}
 0\hspace{1pt}
 0\hspace{1pt}
 0\hspace{1pt}
 0\hspace{1pt}
 0\hspace{1pt} &
 0\hspace{1pt}
 0\hspace{1pt}
 0\hspace{1pt}
 0\hspace{1pt}
 0\hspace{1pt}
 0\hspace{1pt}
 0\hspace{1pt}
 0\hspace{1pt}
 0\hspace{1pt}
 0\hspace{1pt} &
 0\hspace{1pt}
 0\hspace{1pt}
 0\hspace{1pt}
 0\hspace{1pt}
 0\hspace{1pt}
 0\hspace{1pt}
 0\hspace{1pt}
 0\hspace{1pt}
 0\hspace{1pt}
 0\hspace{1pt} &
 0\hspace{1pt}
 0\hspace{1pt}
 0\hspace{1pt}
 0\hspace{1pt}
 0\hspace{1pt}
 0\hspace{1pt}
 0\hspace{1pt}
 0\hspace{1pt} \\   22&
 0\hspace{1pt}  1
 0\hspace{1pt}  1
 0\hspace{1pt}
 0\hspace{1pt}
 0\hspace{1pt}  1  1 &
 0\hspace{1pt}
 0\hspace{1pt}
 0\hspace{1pt}
 0\hspace{1pt}
 0\hspace{1pt}
 0\hspace{1pt}
 0\hspace{1pt}
 0\hspace{1pt}
 0\hspace{1pt}
 0\hspace{1pt} &
 0\hspace{1pt}
 0\hspace{1pt}
 0\hspace{1pt}
 0\hspace{1pt}
 0\hspace{1pt}
 0\hspace{1pt}
 0\hspace{1pt}  1  1  1 & 1  1
 0\hspace{1pt}
 0\hspace{1pt}
 0\hspace{1pt}
 0\hspace{1pt}
 0\hspace{1pt}
 0\hspace{1pt} \\   23&
 0\hspace{1pt}
 0\hspace{1pt}
 0\hspace{1pt}
 0\hspace{1pt}
 0\hspace{1pt}
 0\hspace{1pt}
 0\hspace{1pt}
 0\hspace{1pt}
 0\hspace{1pt} &
 0\hspace{1pt}
 0\hspace{1pt}
 0\hspace{1pt}
 0\hspace{1pt}
 0\hspace{1pt}
 0\hspace{1pt}
 0\hspace{1pt}
 0\hspace{1pt}
 0\hspace{1pt}
 0\hspace{1pt} &
 0\hspace{1pt}
 0\hspace{1pt}
 0\hspace{1pt}
 0\hspace{1pt}
 0\hspace{1pt}
 0\hspace{1pt}
 0\hspace{1pt}
 0\hspace{1pt}
 0\hspace{1pt}
 0\hspace{1pt} &
 0\hspace{1pt}
 0\hspace{1pt}
 0\hspace{1pt}
 0\hspace{1pt}
 0\hspace{1pt}
 0\hspace{1pt}
 0\hspace{1pt}
 0\hspace{1pt} \\   24&
 0\hspace{1pt}
 0\hspace{1pt}
 0\hspace{1pt}
 0\hspace{1pt}
 0\hspace{1pt}
 0\hspace{1pt}
 0\hspace{1pt}
 0\hspace{1pt}
 0\hspace{1pt} &
 0\hspace{1pt}
 0\hspace{1pt}
 0\hspace{1pt}
 0\hspace{1pt}
 0\hspace{1pt}
 0\hspace{1pt}
 0\hspace{1pt}
 0\hspace{1pt}
 0\hspace{1pt}
 0\hspace{1pt} &
 0\hspace{1pt}
 0\hspace{1pt}
 0\hspace{1pt}
 0\hspace{1pt}
 0\hspace{1pt}
 0\hspace{1pt}
 0\hspace{1pt}
 0\hspace{1pt}
 0\hspace{1pt}
 0\hspace{1pt} &
 0\hspace{1pt}
 0\hspace{1pt}
 0\hspace{1pt}
 0\hspace{1pt}
 0\hspace{1pt}
 0\hspace{1pt}
 0\hspace{1pt}
 0\hspace{1pt} \\   25&
 0\hspace{1pt}
 0\hspace{1pt}
 0\hspace{1pt}
 0\hspace{1pt}
 0\hspace{1pt}
 0\hspace{1pt}
 0\hspace{1pt}
 0\hspace{1pt}
 0\hspace{1pt} &
 0\hspace{1pt}
 0\hspace{1pt}
 0\hspace{1pt}
 0\hspace{1pt}
 0\hspace{1pt}
 0\hspace{1pt}
 0\hspace{1pt}
 0\hspace{1pt}
 0\hspace{1pt}
 0\hspace{1pt} &
 0\hspace{1pt}
 0\hspace{1pt}
 0\hspace{1pt}
 0\hspace{1pt}
 0\hspace{1pt}
 0\hspace{1pt}
 0\hspace{1pt}
 0\hspace{1pt}
 0\hspace{1pt}
 0\hspace{1pt} &
 0\hspace{1pt}
 0\hspace{1pt}
 0\hspace{1pt}
 0\hspace{1pt}
 0\hspace{1pt}
 0\hspace{1pt}
 0\hspace{1pt}
 0\hspace{1pt} \\   26&
 0\hspace{1pt}
 0\hspace{1pt}
 0\hspace{1pt}
 0\hspace{1pt}
 0\hspace{1pt}
 0\hspace{1pt}
 0\hspace{1pt}
 0\hspace{1pt}
 0\hspace{1pt} &
 0\hspace{1pt}
 0\hspace{1pt}
 0\hspace{1pt}
 0\hspace{1pt}
 0\hspace{1pt}
 0\hspace{1pt}
 0\hspace{1pt}
 0\hspace{1pt}
 0\hspace{1pt}
 0\hspace{1pt} &
 0\hspace{1pt}
 0\hspace{1pt}
 0\hspace{1pt}
 0\hspace{1pt}
 0\hspace{1pt}
 0\hspace{1pt}
 0\hspace{1pt}
 0\hspace{1pt}
 0\hspace{1pt}
 0\hspace{1pt} &
 0\hspace{1pt}
 0\hspace{1pt}
 0\hspace{1pt}
 0\hspace{1pt}
 0\hspace{1pt}
 0\hspace{1pt}
 0\hspace{1pt}
 0\hspace{1pt} \\   27&
 0\hspace{1pt}
 0\hspace{1pt}
 0\hspace{1pt}
 0\hspace{1pt}
 0\hspace{1pt}
 0\hspace{1pt}
 0\hspace{1pt}
 0\hspace{1pt}  1 &
 0\hspace{1pt}
 0\hspace{1pt}
 0\hspace{1pt}
 0\hspace{1pt}
 0\hspace{1pt}  1
 0\hspace{1pt}
 0\hspace{1pt}
 0\hspace{1pt}
 0\hspace{1pt} &
 0\hspace{1pt}
 0\hspace{1pt}
 0\hspace{1pt}
 0\hspace{1pt}
 0\hspace{1pt}
 0\hspace{1pt}
 0\hspace{1pt}
 0\hspace{1pt}  1
 0\hspace{1pt} &
 0\hspace{1pt}
 0\hspace{1pt}
 0\hspace{1pt}
 0\hspace{1pt}
 0\hspace{1pt}
 0\hspace{1pt}
 0\hspace{1pt}
 0\hspace{1pt} \\   28&
 0\hspace{1pt}
 0\hspace{1pt}
 0\hspace{1pt}
 0\hspace{1pt}
 0\hspace{1pt}
 0\hspace{1pt}
 0\hspace{1pt}
 0\hspace{1pt}
 0\hspace{1pt} &
 0\hspace{1pt}
 0\hspace{1pt}
 0\hspace{1pt}
 0\hspace{1pt}
 0\hspace{1pt}
 0\hspace{1pt}
 0\hspace{1pt}
 0\hspace{1pt}
 0\hspace{1pt}
 0\hspace{1pt} &
 0\hspace{1pt}
 0\hspace{1pt}
 0\hspace{1pt}
 0\hspace{1pt}
 0\hspace{1pt}
 0\hspace{1pt}
 0\hspace{1pt}
 0\hspace{1pt}  0
 0\hspace{1pt} &
 0\hspace{1pt}
 0\hspace{1pt}
 0\hspace{1pt}
 0\hspace{1pt}
 0\hspace{1pt}
 0\hspace{1pt}
 0\hspace{1pt}
 0\hspace{1pt} \\   29&
 0\hspace{1pt}
 0\hspace{1pt}
 0\hspace{1pt}
 0\hspace{1pt}
 0\hspace{1pt}
 0\hspace{1pt}
 0\hspace{1pt}
 0\hspace{1pt}
 0\hspace{1pt} &
 0\hspace{1pt}
 0\hspace{1pt}
 0\hspace{1pt}
 0\hspace{1pt}
 0\hspace{1pt}
 0\hspace{1pt}
 0\hspace{1pt}
 0\hspace{1pt}
 0\hspace{1pt}
 0\hspace{1pt} &
 0\hspace{1pt}
 0\hspace{1pt}
 0\hspace{1pt}
 0\hspace{1pt}
 0\hspace{1pt}
 0\hspace{1pt}
 0\hspace{1pt}
 0\hspace{1pt}
 0\hspace{1pt}  0 &
 0\hspace{1pt}
 0\hspace{1pt}
 0\hspace{1pt}
 0\hspace{1pt}
 0\hspace{1pt}
 0\hspace{1pt}
 0\hspace{1pt}
 0\hspace{1pt} \\   30&
 0\hspace{1pt}
 0\hspace{1pt}
 0\hspace{1pt}
 0\hspace{1pt}
 0\hspace{1pt}
 0\hspace{1pt}
 0\hspace{1pt}
 0\hspace{1pt}
 0\hspace{1pt} &
 0\hspace{1pt}
 0\hspace{1pt}
 0\hspace{1pt}
 0\hspace{1pt}
 0\hspace{1pt}
 0\hspace{1pt}
 0\hspace{1pt}
 0\hspace{1pt}
 0\hspace{1pt}
 0\hspace{1pt} &
 0\hspace{1pt}
 0\hspace{1pt}
 0\hspace{1pt}
 0\hspace{1pt}
 0\hspace{1pt}
 0\hspace{1pt}
 0\hspace{1pt}
 0\hspace{1pt}
 0\hspace{1pt}
 0\hspace{1pt} & 0
 0\hspace{1pt}
 0\hspace{1pt}
 0\hspace{1pt}
 0\hspace{1pt}
 0\hspace{1pt}
 0\hspace{1pt}
 0\hspace{1pt} \\   31&
 0\hspace{1pt}
 0\hspace{1pt}
 0\hspace{1pt}
 0\hspace{1pt}
 0\hspace{1pt}
 0\hspace{1pt}
 0\hspace{1pt}
 0\hspace{1pt}
 0\hspace{1pt} &
 0\hspace{1pt}
 0\hspace{1pt}
 0\hspace{1pt}
 0\hspace{1pt}
 0\hspace{1pt}
 0\hspace{1pt}
 0\hspace{1pt}
 0\hspace{1pt}
 0\hspace{1pt}
 0\hspace{1pt} &
 0\hspace{1pt}
 0\hspace{1pt}
 0\hspace{1pt}
 0\hspace{1pt}
 0\hspace{1pt}
 0\hspace{1pt}
 0\hspace{1pt}
 0\hspace{1pt}  1
 0\hspace{1pt} &
 0\hspace{1pt}
 0\hspace{1pt}
 0\hspace{1pt}
 0\hspace{1pt}
 0\hspace{1pt}
 0\hspace{1pt}
 0\hspace{1pt}
 0\hspace{1pt} \\   32&
 0\hspace{1pt}
 0\hspace{1pt}
 0\hspace{1pt}
 0\hspace{1pt}
 0\hspace{1pt}
 0\hspace{1pt}
 0\hspace{1pt}
 0\hspace{1pt}
 0\hspace{1pt} &
 0\hspace{1pt}
 0\hspace{1pt}
 0\hspace{1pt}
 0\hspace{1pt}
 0\hspace{1pt}
 0\hspace{1pt}
 0\hspace{1pt}
 0\hspace{1pt}
 0\hspace{1pt}
 0\hspace{1pt} &
 0\hspace{1pt}
 0\hspace{1pt}
 0\hspace{1pt}
 0\hspace{1pt}
 0\hspace{1pt}
 0\hspace{1pt}
 0\hspace{1pt}
 0\hspace{1pt}
 0\hspace{1pt}
 0\hspace{1pt} &
 0\hspace{1pt}
 0\hspace{1pt}  0
 0\hspace{1pt}
 0\hspace{1pt}
 0\hspace{1pt}
 0\hspace{1pt}
 0\hspace{1pt} \\   33&
 0\hspace{1pt}
 0\hspace{1pt}
 0\hspace{1pt}
 0\hspace{1pt}
 0\hspace{1pt}
 0\hspace{1pt}
 0\hspace{1pt}
 0\hspace{1pt}
 0\hspace{1pt} &
 0\hspace{1pt}
 0\hspace{1pt}
 0\hspace{1pt}
 0\hspace{1pt}
 0\hspace{1pt}
 0\hspace{1pt}
 0\hspace{1pt}
 0\hspace{1pt}
 0\hspace{1pt}
 0\hspace{1pt} &
 0\hspace{1pt}
 0\hspace{1pt}
 0\hspace{1pt}
 0\hspace{1pt}
 0\hspace{1pt}
 0\hspace{1pt}
 0\hspace{1pt}
 0\hspace{1pt}
 0\hspace{1pt}
 0\hspace{1pt} &
 0\hspace{1pt}
 0\hspace{1pt}
 0\hspace{1pt}
 0\hspace{1pt}
 0\hspace{1pt}
 0\hspace{1pt}
 0\hspace{1pt}
 0\hspace{1pt} \\   34&
 0\hspace{1pt}
 0\hspace{1pt}
 0\hspace{1pt}
 0\hspace{1pt}
 0\hspace{1pt}
 0\hspace{1pt}
 0\hspace{1pt}
 0\hspace{1pt}
 0\hspace{1pt} &
 0\hspace{1pt}
 0\hspace{1pt}
 0\hspace{1pt}
 0\hspace{1pt}
 0\hspace{1pt}
 0\hspace{1pt}
 0\hspace{1pt}
 0\hspace{1pt}
 0\hspace{1pt}
 0\hspace{1pt} &
 0\hspace{1pt}
 0\hspace{1pt}
 0\hspace{1pt}
 0\hspace{1pt}
 0\hspace{1pt}
 0\hspace{1pt}
 0\hspace{1pt}
 0\hspace{1pt}
 0\hspace{1pt}
 0\hspace{1pt} &
 0\hspace{1pt}
 0\hspace{1pt}
 0\hspace{1pt}
 0\hspace{1pt}
 0\hspace{1pt}
 0\hspace{1pt}
 0\hspace{1pt}
 0\hspace{1pt} \\   35&
 0\hspace{1pt}
 0\hspace{1pt}
 0\hspace{1pt}
 0\hspace{1pt}
 0\hspace{1pt}
 0\hspace{1pt}
 0\hspace{1pt}
 0\hspace{1pt}
 0\hspace{1pt} &
 0\hspace{1pt}
 0\hspace{1pt}
 0\hspace{1pt}
 0\hspace{1pt}
 0\hspace{1pt}
 0\hspace{1pt}
 0\hspace{1pt}
 0\hspace{1pt}
 0\hspace{1pt}
 0\hspace{1pt} &
 0\hspace{1pt}
 0\hspace{1pt}
 0\hspace{1pt}  1
 0\hspace{1pt}
 0\hspace{1pt}
 0\hspace{1pt}
 0\hspace{1pt}
 0\hspace{1pt}
 0\hspace{1pt} &
 0\hspace{1pt}
 0\hspace{1pt}
 0\hspace{1pt}
 0\hspace{1pt}
 0\hspace{1pt}
 0\hspace{1pt}
 0\hspace{1pt}
 0\hspace{1pt} \\   36&
 0\hspace{1pt}
 0\hspace{1pt}
 0\hspace{1pt}
 0\hspace{1pt}
 0\hspace{1pt}
 0\hspace{1pt}
 0\hspace{1pt}
 0\hspace{1pt}
 0\hspace{1pt} &
 0\hspace{1pt}
 0\hspace{1pt}
 0\hspace{1pt}
 0\hspace{1pt}
 0\hspace{1pt}
 0\hspace{1pt}
 0\hspace{1pt}
 0\hspace{1pt}
 0\hspace{1pt}
 0\hspace{1pt} &
 0\hspace{1pt}
 0\hspace{1pt}
 0\hspace{1pt}
 0\hspace{1pt}
 0\hspace{1pt}
 0\hspace{1pt}
 0\hspace{1pt}
 0\hspace{1pt}
 0\hspace{1pt}
 0\hspace{1pt} &
 0\hspace{1pt}
 0\hspace{1pt}
 0\hspace{1pt}
 0\hspace{1pt}
 0\hspace{1pt}
 0\hspace{1pt}  0
 0\hspace{1pt} \\   37&
 0\hspace{1pt}
 0\hspace{1pt}
 0\hspace{1pt}
 0\hspace{1pt}
 0\hspace{1pt}
 0\hspace{1pt}
 0\hspace{1pt}
 0\hspace{1pt}
 0\hspace{1pt} &
 0\hspace{1pt}
 0\hspace{1pt}
 0\hspace{1pt}
 0\hspace{1pt}
 0\hspace{1pt}
 0\hspace{1pt}
 0\hspace{1pt}
 0\hspace{1pt}
 0\hspace{1pt}
 0\hspace{1pt} &
 0\hspace{1pt}
 0\hspace{1pt}
 0\hspace{1pt}
 0\hspace{1pt}
 0\hspace{1pt}
 0\hspace{1pt}
 0\hspace{1pt}
 0\hspace{1pt}
 0\hspace{1pt}
 0\hspace{1pt} &
 0\hspace{1pt}
 0\hspace{1pt}
 0\hspace{1pt}
 0\hspace{1pt}
 0\hspace{1pt}
 0\hspace{1pt}
 0\hspace{1pt}
 0\hspace{1pt} \\  
 \hline \end{tabular} \caption{Matrix $C_0$}\end{table}

\newpage
\begin{table}\label{table:D} \begin{tabular}{|l|l|l|l|l|} \hline $D_0$&{\small  38 \hspace{4pt} $\cdots$\hspace{5pt} 43 \hspace{4pt}$\cdots$ \hspace{5pt}47} &{\small 48 \hspace{3pt}  $\cdots$ \hspace{3pt}53 \hspace{3pt}$\cdots$\hspace{3pt} 57 }& {\small 58 \hspace{4pt} $\cdots$ 63 \hspace{4pt} $\cdots$ 67 }&{\small 68 \hspace{4pt} $\cdots$ 73 \hspace{4pt} $\cdots$ 77 }\\ \hline
38& 0  1  1  1    0 
  0   1  1
  0   1 &
  0 
  0 
  0 
  0 
  0 
  0 
  0 
  0 
  0 
  0  &
  0 
  0 
  0   1
  0 
  0 
  0 
  0 
  0   1 &
  0   1  1
  0 
  0 
  0 
  0 
  0   1
  0  \\ \hline 39& 1  0  1
  0 
  0 
  0 
  0   1
  0 
  0  &
  0 
  0 
  0 
  0 
  0 
  0 
  0 
  0 
  0 
  0  &
  0 
  0 
  0 
  0 
  0 
  0 
  0 
  0 
  0 
  0  &
  0   1
  0 
  0 
  0 
  0 
  0 
  0 
  0 
  0  \\ \hline 40& 1  1  0  1
  0 
  0   1  1
  0 
  0  &
  0 
  0 
  0 
  0 
  0 
  0 
  0 
  0 
  0 
  0  &
  0 
  0 
  0   1
  0 
  0 
  0 
  0 
  0 
  0  &
  0 
  0 
  0 
  0 
  0 
  0 
  0 
  0 
  0   1 \\ \hline 41&
  0 
  0 
  0   0
  0   1
  0 
  0 
  0 
  0  &
  0 
  0 
  0 
  0 
  0 
  0 
  0 
  0 
  0 
  0  &
  0 
  0 
  0 
  0 
  0 
  0 
  0 
  0 
  0 
  0  &
  0 
  0 
  0 
  0 
  0 
  0 
  0 
  0 
  0   1 \\ \hline 42&
  0 
  0 
  0   1
  0   1
  0 
  0 
  0 
  0  &
  0 
  0 
  0 
  0 
  0 
  0 
  0 
  0 
  0 
  0  &
  0 
  0 
  0 
  0 
  0 
  0 
  0 
  0 
  0   1 &
  0 
  0 
  0 
  0 
  0 
  0 
  0 
  0 
  0   1 \\ \hline 43& 1
  0 
  0   1
  0   0
  0 
  0 
  0 
  0  &
  0 
  0 
  0 
  0 
  0 
  0 
  0 
  0 
  0 
  0  &
  0 
  0   1
  0 
  0 
  0 
  0 
  0 
  0 
  0  &
  0 
  0 
  0 
  0 
  0 
  0 
  0 
  0 
  0 
  0  \\ \hline 44&
  0 
  0 
  0   1
  0 
  0   1  0
  0 
  0  &
  0 
  0   1
  0 
  0 
  0 
  0 
  0 
  0 
  0  &
  0 
  0 
  0 
  0 
  0 
  0 
  0 
  0 
  0 
  0  & 1
  0 
  0 
  0 
  0 
  0 
  0 
  0   1  1 \\ \hline 45& 1
  0 
  0 
  0 
  0 
  0 
  0   0
  0 
  0  &
  0 
  0 
  0 
  0 
  0 
  0 
  0 
  0 
  0   1 &
  0   1
  0 
  0 
  0 
  0 
  0 
  0 
  0 
  0  &
  0 
  0 
  0 
  0 
  0 
  0   1
  0   1
  0  \\ \hline 46& 1
  0 
  0   1
  0 
  0 
  0 
  0 
  0 
  0  &
  0 
  0 
  0 
  0 
  0 
  0 
  0 
  0 
  0 
  0  &
  0 
  0 
  0 
  0 
  0 
  0 
  0 
  0 
  0 
  0  &
  0 
  0 
  0 
  0 
  0 
  0 
  0 
  0 
  0 
  0  \\ \hline 47&
  0 
  0   1
  0 
  0 
  0 
  0 
  0 
  0 
  0  &
  0 
  0 
  0 
  0 
  0 
  0 
  0 
  0 
  0 
  0  &
  0 
  0 
  0 
  0 
  0 
  0 
  0 
  0 
  0 
  0  &
  0 
  0 
  0 
  0 
  0 
  0 
  0 
  0 
  0 
  0  \\ \hline 48&
  0 
  0 
  0 
  0 
  0 
  0 
  0 
  0 
  0   1 &
  0 
  0 
  0 
  0 
  0 
  0 
  0 
  0 
  0 
  0  &
  0 
  0 
  0 
  0 
  0 
  0 
  0 
  0 
  0 
  0  &
  0 
  0 
  0 
  0 
  0 
  0 
  0 
  0 
  0 
  0  \\ \hline 49&
  0 
  0 
  0 
  0 
  0 
  0 
  0 
  0 
  0 
  0  &
  0 
  0 
  0 
  0 
  0 
  0 
  0 
  0 
  0 
  0  &
  0 
  0 
  0 
  0 
  0 
  0 
  0 
  0 
  0 
  0  &
  0 
  0 
  0 
  0 
  0 
  0 
  0 
  0 
  0 
  0  \\ \hline 50& 1
  0 
  0 
  0 
  0 
  0 
  0 
  0 
  0   1 &
  0 
  0   0
  0 
  0 
  0 
  0 
  0 
  0 
  0  &
  0 
  0 
  0   1
  0 
  0   1
  0 
  0 
  0  &
  0 
  0 
  0 
  0 
  0 
  0 
  0 
  0   1
  0  \\ \hline 51&
  0 
  0 
  0 
  0 
  0 
  0 
  0 
  0 
  0 
  0  &
  0 
  0 
  0 
  0 
  0 
  0 
  0 
  0 
  0 
  0  &
  0 
  0 
  0   1
  0 
  0 
  0 
  0 
  0 
  0  &
  0 
  0 
  0 
  0 
  0 
  0 
  0 
  0 
  0 
  0  \\ \hline 52&
  0 
  0 
  0 
  0 
  0 
  0 
  0 
  0 
  0 
  0  &
  0 
  0 
  0 
  0 
  0 
  0 
  0 
  0 
  0 
  0  &
  0 
  0 
  0 
  0 
  0 
  0 
  0 
  0 
  0 
  0  &
  0 
  0 
  0 
  0 
  0 
  0 
  0 
  0 
  0 
  0  \\ \hline 53&
  0 
  0   1
  0 
  0 
  0 
  0   1
  0 
  0  &
  0 
  0 
  0 
  0 
  0   1
  0 
  0 
  0   1 &
  0 
  0 
  0 
  0 
  0 
  0 
  0 
  0 
  0 
  0  &
  0 
  0 
  0 
  0 
  0 
  0 
  0 
  0   1
  0  \\ \hline 54&
  0 
  0 
  0 
  0 
  0 
  0 
  0 
  0 
  0 
  0  &
  0 
  0 
  0 
  0 
  0 
  0 
  0 
  0 
  0 
  0  &
  0 
  0 
  0 
  0 
  0 
  0 
  0 
  0 
  0 
  0  &
  0 
  0 
  0 
  0 
  0 
  0 
  0 
  0 
  0 
  0  \\ \hline 55&
  0 
  0 
  0 
  0 
  0 
  0 
  0 
  0 
  0 
  0  &
  0 
  0 
  0 
  0 
  0 
  0 
  0 
  0 
  0 
  0  &
  0 
  0 
  0 
  0 
  0 
  0 
  0 
  0 
  0 
  0  &
  0 
  0 
  0 
  0 
  0 
  0 
  0 
  0 
  0 
  0  \\ \hline 56&
  0 
  0 
  0 
  0 
  0 
  0 
  0 
  0 
  0 
  0  &
  0 
  0 
  0 
  0 
  0 
  0 
  0 
  0 
  0 
  0  &
  0 
  0 
  0 
  0 
  0 
  0 
  0 
  0 
  0 
  0  &
  0 
  0 
  0 
  0 
  0 
  0 
  0 
  0 
  0 
  0  \\ \hline 57&
  0 
  0 
  0 
  0 
  0 
  0 
  0 
  0 
  0 
  0  &
  0 
  0 
  0 
  0 
  0 
  0 
  0 
  0 
  0 
  0  &
  0 
  0 
  0 
  0 
  0 
  0 
  0 
  0 
  0 
  0  &
  0 
  0 
  0 
  0 
  0 
  0 
  0 
  0 
  0 
  0  \\ \hline 58&
  0 
  0 
  0 
  0 
  0 
  0 
  0 
  0 
  0 
  0  &
  0 
  0 
  0 
  0 
  0 
  0 
  0 
  0 
  0 
  0  & 0
  0 
  0 
  0 
  0 
  0 
  0 
  0 
  0 
  0  &
  0 
  0 
  0 
  0 
  0 
  0 
  0 
  0 
  0 
  0  \\ \hline 59&
  0 
  0 
  0 
  0 
  0 
  0 
  0 
  0 
  0 
  0  &
  0 
  0 
  0 
  0 
  0 
  0 
  0 
  0 
  0 
  0  &
  0 
  0 
  0 
  0 
  0 
  0 
  0 
  0 
  0 
  0  &
  0 
  0 
  0 
  0 
  0 
  0 
  0 
  0 
  0 
  0  \\ \hline 60& 1
  0 
  0 
  0 
  0 
  0 
  0 
  0 
  0 
  0  &
  0 
  0 
  0 
  0 
  0 
  0 
  0 
  0 
  0 
  0  &
  0 
  0   0
  0 
  0 
  0 
  0 
  0 
  0 
  0  &
  0 
  0 
  0 
  0 
  0 
  0 
  0 
  0 
  0 
  0  \\ \hline 61&
  0 
  0 
  0 
  0 
  0 
  0 
  0 
  0 
  0 
  0  &
  0 
  0 
  0 
  0 
  0 
  0 
  0 
  0 
  0 
  0  &
  0 
  0 
  0 
  0 
  0 
  0 
  0 
  0 
  0 
  0  &
  0 
  0 
  0 
  0 
  0 
  0 
  0 
  0 
  0 
  0  \\ \hline 62&
  0 
  0 
  0 
  0 
  0 
  0 
  0 
  0 
  0 
  0  &
  0 
  0 
  0 
  0 
  0 
  0 
  0 
  0 
  0 
  0  &
  0 
  0 
  0 
  0 
  0 
  0 
  0 
  0 
  0 
  0  &
  0 
  0 
  0 
  0 
  0 
  0 
  0 
  0 
  0 
  0  \\ \hline 63&
  0 
  0 
  0 
  0 
  0 
  0 
  0 
  0 
  0 
  0  &
  0 
  0 
  0 
  0 
  0 
  0 
  0 
  0 
  0 
  0  &
  0 
  0 
  0 
  0 
  0 
  0 
  0 
  0 
  0 
  0  &
  0 
  0 
  0 
  0 
  0 
  0 
  0 
  0 
  0 
  0  \\ \hline 64&
  0 
  0 
  0 
  0 
  0 
  0 
  0 
  0 
  0 
  0  &
  0 
  0 
  0 
  0 
  0 
  0 
  0 
  0 
  0 
  0  &
  0 
  0 
  0 
  0 
  0 
  0 
  0 
  0 
  0 
  0  &
  0 
  0 
  0 
  0 
  0 
  0 
  0 
  0 
  0 
  0  \\ \hline 65& 1
  0 
  0 
  0 
  0 
  0 
  0   1
  0 
  0  &
  0 
  0 
  0 
  0 
  0   1
  0 
  0 
  0 
  0  &
  0   1
  0 
  0 
  0 
  0 
  0 
  0   1
  0  &
  0 
  0 
  0 
  0 
  0 
  0 
  0 
  0 
  0 
  0  \\ \hline 66&
  0 
  0 
  0 
  0 
  0 
  0 
  0 
  0 
  0 
  0  &
  0 
  0 
  0 
  0 
  0 
  0 
  0 
  0 
  0 
  0  &
  0 
  0 
  0 
  0 
  0 
  0 
  0 
  0   0.
  0  &
  0 
  0 
  0 
  0 
  0 
  0 
  0 
  0 
  0 
  0  \\ \hline 67&
  0 
  0 
  0 
  0 
  0 
  0 
  0 
  0 
  0 
  0  &
  0 
  0 
  0 
  0 
  0 
  0 
  0 
  0 
  0 
  0  &
  0 
  0 
  0 
  0 
  0 
  0 
  0 
  0 
  0 
  0  &
  0 
  0 
  0 
  0 
  0 
  0 
  0 
  0 
  0 
  0  \\ \hline 68&
  0 
  0 
  0 
  0 
  0 
  0 
  0 
  0 
  0 
  0  &
  0 
  0 
  0 
  0 
  0 
  0 
  0 
  0 
  0 
  0  &
  0 
  0 
  0 
  0 
  0 
  0 
  0 
  0 
  0 
  0  &
  0 
  0 
  0 
  0 
  0 
  0 
  0 
  0 
  0 
  0  \\ \hline 69&
  0 
  0 
  0 
  0 
  0 
  0 
  0 
  0 
  0 
  0  &
  0 
  0 
  0 
  0 
  0 
  0 
  0 
  0 
  0 
  0  &
  0 
  0 
  0 
  0 
  0 
  0 
  0 
  0 
  0 
  0  &
  0 
  0 
  0 
  0 
  0 
  0 
  0 
  0 
  0 
  0  \\ \hline 70&
  0 
  0 
  0 
  0 
  0 
  0 
  0 
  0 
  0 
  0  &
  0 
  0 
  0 
  0 
  0 
  0 
  0 
  0 
  0 
  0  &
  0 
  0 
  0 
  0 
  0 
  0 
  0 
  0 
  0 
  0  &
  0 
  0 
  0 
  0 
  0 
  0 
  0 
  0 
  0 
  0  \\ \hline 71&
  0 
  0 
  0 
  0 
  0 
  0 
  0 
  0 
  0 
  0  &
  0 
  0 
  0 
  0 
  0 
  0 
  0 
  0 
  0 
  0  &
  0 
  0 
  0 
  0 
  0 
  0 
  0 
  0 
  0 
  0  &
  0 
  0 
  0   0.
  0 
  0 
  0 
  0   1
  0  \\ \hline 72&
  0 
  0 
  0 
  0 
  0 
  0 
  0 
  0 
  0 
  0  &
  0 
  0 
  0 
  0 
  0 
  0 
  0 
  0 
  0 
  0  &
  0 
  0 
  0 
  0 
  0 
  0 
  0 
  0 
  0 
  0  &
  0 
  0 
  0 
  0 
  0 
  0 
  0 
  0 
  0 
  0  \\ \hline 73&
  0 
  0 
  0 
  0 
  0 
  0 
  0 
  0 
  0 
  0  &
  0 
  0 
  0 
  0 
  0 
  0 
  0 
  0 
  0 
  0  &
  0 
  0 
  0 
  0 
  0 
  0 
  0 
  0 
  0 
  0  &
  0 
  0 
  0 
  0 
  0 
  0 
  0 
  0 
  0 
  0  \\ \hline 74&
  0 
  0 
  0 
  0 
  0 
  0 
  0 
  0 
  0 
  0  &
  0 
  0 
  0 
  0 
  0 
  0 
  0 
  0 
  0 
  0  &
  0 
  0 
  0 
  0 
  0 
  0 
  0 
  0 
  0 
  0  &
  0 
  0 
  0 
  0 
  0 
  0 
  0 
  0   1
  0  \\ \hline 75&
  0 
  0 
  0 
  0 
  0 
  0 
  0 
  0 
  0 
  0  &
  0 
  0 
  0 
  0 
  0 
  0 
  0 
  0 
  0 
  0  &
  0 
  0 
  0 
  0 
  0 
  0 
  0 
  0 
  0 
  0  &
  0 
  0 
  0 
  0 
  0 
  0 
  0 
  0 
  0 
  0  \\ \hline 76& 1  1
  0 
  0 
  0 
  0   1  1
  0 
  0  &
  0 
  0   1
  0 
  0 
  0   1
  0 
  0 
  0  &
  0 
  0 
  0 
  0 
  0 
  0 
  0 
  0 
  0 
  0  &
  0 
  0 
  0 
  0 
  0 
  0 
  0 
  0   0.
  0  \\ \hline 77&
  0 
  0 
  0   1
  0 
  0 
  0 
  0 
  0 
  0  &
  0 
  0 
  0 
  0 
  0 
  0 
  0 
  0 
  0 
  0  &
  0 
  0 
  0 
  0 
  0 
  0 
  0 
  0 
  0 
  0  &
  0 
  0 
  0 
  0 
  0 
  0 
  0 
  0 
  0   0. \\ \hline
\end{tabular}\caption{Matrix D} \end{table}
\newpage
\newpage
\begin{table}\label{table:A} \begin{tabular}{|l|l|l|l|l|} \hline $A$&{\small  38 \hspace{4pt} $\cdots$\hspace{5pt} 43 \hspace{4pt}$\cdots$ \hspace{5pt}47} &{\small 48 \hspace{3pt}  $\cdots$ \hspace{3pt}53 \hspace{3pt}$\cdots$\hspace{3pt} 57 }& {\small 58 \hspace{4pt} $\cdots$ 63 \hspace{4pt} $\cdots$ 67 }&{\small 68 \hspace{4pt} $\cdots$ 73 \hspace{4pt} $\cdots$ 77 }\\ \hline
 1& 1  1  1  1  1  1  1  1    0     0  &   0     0     0     0   1  1    0     0   1    0  &   0     0     0     0   1    0   1    0     0   1 & 1    0     0     0     0     0     0     0   1  1 \\ \hline
 2& 1  1  1  1  1    0   1  1  1  1 &   0   1  1  1    0   1    0     0     0     0  &   0   1    0   1  1    0     0     0     0   1 & 1    0     0   1    0     0     0     0   1  1 \\ \hline
 3& 1    0   1  1    0     0   1  1    0     0  &   0     0     0     0   1  1  1  1  1  1 &   0     0     0     0     0     0     0     0     0     0  &   0     0     0     0     0     0     0     0   1    0  \\ \hline
 4&   0     0   1  1    0     0     0     0     0     0  &   0     0     0     0     0     0     0     0     0     0  & 1    0     0     0     0     0     0     0     0     0  &   0     0     0     0     0     0     0     0     0     0  \\ \hline
 5& 1    0     0     0     0     0     0     0   1    0  &   0     0     0     0     0     0     0     0     0     0  &   0   1    0     0     0     0     0     0     0     0  &   0     0     0     0     0     0     0     0     0     0  \\ \hline
 6&   0     0     0     0     0     0     0     0   1    0  &   0     0     0   1    0     0     0     0     0     0  &   0     0     0     0     0     0     0     0     0     0  &   0     0     0     0     0     0     0     0     0     0  \\ \hline
 7& 1    0     0     0     0     0     0   1    0     0  &   0     0     0   1    0     0     0     0     0     0  &   0     0     0   1  1  1    0     0     0     0  &   0     0     0     0     0   1    0     0     0     0  \\ \hline
 8&   0     0   1    0     0     0     0     0     0     0  &   0     0     0     0     0     0     0     0     0     0  &   0   1    0     0     0     0     0     0     0     0  &   0     0     0     0     0     0     0     0     0     0  \\ \hline
 9&   0     0     0     0     0     0     0   1    0     0  &   0     0     0     0     0     0     0     0     0     0  &   0     0     0     0     0     0   1    0     0     0  &   0     0     0     0     0     0     0     0     0     0  \\ \hline
10& 1    0 
  0 
  0 
  0 
  0 
  0 
  0 
  0 
  0  &
  0 
  0 
  0 
  0 
  0 
  0 
  0 
  0 
  0 
  0  &
  0 
  0 
  0 
  0 
  0 
  0 
  0   1  1
  0  &
  0 
  0 
  0 
  0 
  0 
  0   1  1
  0 
  0  \\ \hline 11&
  0 
  0 
  0   1
  0 
  0 
  0   1
  0 
  0  &
  0 
  0   1
  0 
  0 
  0 
  0 
  0 
  0 
  0  &
  0 
  0 
  0 
  0 
  0 
  0 
  0 
  0 
  0 
  0  & 1
  0 
  0 
  0 
  0 
  0 
  0 
  0   1
  0  \\ \hline 12&
  0 
  0 
  0 
  0 
  0 
  0 
  0 
  0 
  0 
  0  &
  0 
  0 
  0 
  0 
  0 
  0 
  0 
  0 
  0 
  0  &
  0 
  0 
  0 
  0 
  0 
  0 
  0 
  0 
  0 
  0  &
  0 
  0 
  0 
  0 
  0 
  0 
  0 
  0 
  0 
  0  \\ \hline 13&
  0 
  0 
  0 
  0 
  0 
  0 
  0 
  0 
  0 
  0  &
  0 
  0 
  0 
  0 
  0 
  0 
  0 
  0 
  0 
  0  &
  0 
  0 
  0 
  0 
  0 
  0 
  0 
  0 
  0 
  0  &
  0 
  0 
  0 
  0 
  0 
  0 
  0 
  0 
  0 
  0  \\ \hline 14&
  0 
  0 
  0 
  0 
  0 
  0 
  0 
  0 
  0 
  0  &
  0 
  0 
  0 
  0 
  0 
  0 
  0 
  0 
  0 
  0  &
  0 
  0 
  0 
  0 
  0 
  0 
  0 
  0 
  0 
  0  &
  0 
  0 
  0 
  0 
  0 
  0 
  0 
  0 
  0 
  0  \\ \hline 15&
  0 
  0   1
  0 
  0 
  0 
  0 
  0 
  0 
  0  &
  0 
  0 
  0 
  0 
  0 
  0 
  0 
  0 
  0 
  0  &
  0 
  0 
  0 
  0 
  0 
  0 
  0 
  0 
  0 
  0  &
  0 
  0 
  0 
  0 
  0 
  0 
  0 
  0 
  0 
  0  \\ \hline 16&
  0 
  0   1
  0 
  0 
  0 
  0 
  0 
  0 
  0  &
  0 
  0 
  0 
  0 
  0 
  0 
  0 
  0 
  0 
  0  &
  0 
  0 
  0 
  0 
  0 
  0 
  0 
  0 
  0 
  0  &
  0 
  0 
  0 
  0 
  0 
  0 
  0 
  0 
  0 
  0  \\ \hline 17&
  0 
  0 
  0 
  0 
  0 
  0 
  0 
  0 
  0 
  0  &
  0 
  0 
  0 
  0 
  0 
  0 
  0 
  0 
  0 
  0  &
  0 
  0 
  0 
  0 
  0 
  0 
  0 
  0 
  0 
  0  &
  0 
  0 
  0 
  0 
  0 
  0 
  0 
  0 
  0 
  0  \\ \hline 18&
  0 
  0 
  0 
  0 
  0 
  0 
  0 
  0 
  0 
  0  &
  0 
  0 
  0 
  0 
  0 
  0 
  0 
  0 
  0 
  0  &
  0 
  0 
  0 
  0 
  0 
  0 
  0 
  0 
  0 
  0  &
  0 
  0 
  0 
  0 
  0 
  0 
  0 
  0 
  0 
  0  \\ \hline 19&
  0 
  0 
  0 
  0 
  0 
  0 
  0 
  0 
  0 
  0  &
  0 
  0 
  0 
  0 
  0 
  0 
  0 
  0 
  0 
  0  &
  0 
  0 
  0 
  0 
  0 
  0 
  0 
  0 
  0 
  0  &
  0 
  0 
  0 
  0 
  0 
  0 
  0 
  0 
  0 
  0  \\ \hline 20&
  0 
  0   1
  0 
  0 
  0   1
  0 
  0 
  0  &
  0 
  0 
  0 
  0 
  0 
  0 
  0 
  0 
  0 
  0  &
  0 
  0 
  0 
  0 
  0 
  0 
  0 
  0 
  0 
  0  &
  0 
  0 
  0 
  0 
  0 
  0 
  0 
  0 
  0 
  0  \\ \hline 21&
  0 
  0 
  0 
  0 
  0 
  0 
  0 
  0 
  0 
  0  &
  0 
  0 
  0 
  0 
  0 
  0 
  0 
  0 
  0 
  0  &
  0 
  0 
  0 
  0 
  0 
  0 
  0 
  0 
  0 
  0  &
  0 
  0 
  0 
  0 
  0 
  0 
  0 
  0 
  0 
  0  \\ \hline 22&
  0 
  0   1
  0 
  0 
  0 
  0 
  0 
  0 
  0  &
  0 
  0 
  0 
  0 
  0 
  0 
  0 
  0 
  0 
  0  &
  0   1
  0 
  0 
  0 
  0 
  0 
  0 
  0 
  0  &
  0 
  0 
  0 
  0 
  0 
  0 
  0 
  0 
  0 
  0  \\ \hline 23&
  0 
  0 
  0 
  0 
  0 
  0 
  0 
  0 
  0 
  0  &
  0 
  0 
  0 
  0 
  0 
  0 
  0 
  0 
  0 
  0  &
  0 
  0 
  0 
  0 
  0 
  0 
  0 
  0 
  0 
  0  &
  0 
  0 
  0 
  0 
  0 
  0 
  0 
  0 
  0 
  0  \\ \hline 24&
  0 
  0 
  0 
  0 
  0 
  0 
  0 
  0 
  0 
  0  &
  0 
  0 
  0 
  0 
  0 
  0 
  0 
  0 
  0 
  0  &
  0 
  0 
  0 
  0 
  0 
  0 
  0 
  0 
  0 
  0  &
  0 
  0 
  0 
  0 
  0 
  0 
  0 
  0 
  0 
  0  \\ \hline 25&
  0 
  0 
  0 
  0 
  0 
  0 
  0 
  0 
  0 
  0  &
  0 
  0 
  0 
  0 
  0 
  0 
  0 
  0 
  0 
  0  &
  0 
  0 
  0 
  0 
  0 
  0 
  0 
  0 
  0 
  0  &
  0 
  0 
  0 
  0 
  0 
  0 
  0 
  0 
  0 
  0  \\ \hline 26&
  0 
  0 
  0 
  0 
  0 
  0 
  0 
  0 
  0 
  0  &
  0 
  0 
  0 
  0 
  0 
  0 
  0 
  0 
  0 
  0  &
  0 
  0 
  0 
  0 
  0 
  0 
  0 
  0 
  0 
  0  &
  0 
  0 
  0 
  0 
  0 
  0 
  0 
  0 
  0 
  0  \\ \hline 27&
  0 
  0 
  0 
  0 
  0 
  0 
  0 
  0 
  0 
  0  &
  0 
  0 
  0 
  0 
  0 
  0 
  0 
  0 
  0 
  0  &
  0 
  0 
  0 
  0 
  0 
  0 
  0 
  0 
  0 
  0  &
  0 
  0 
  0 
  0 
  0 
  0 
  0 
  0 
  0   1 \\ \hline 28&
  0 
  0 
  0 
  0 
  0 
  0 
  0 
  0 
  0 
  0  &
  0 
  0 
  0 
  0 
  0 
  0 
  0 
  0 
  0 
  0  &
  0 
  0 
  0 
  0 
  0 
  0 
  0 
  0 
  0 
  0  &
  0 
  0 
  0 
  0 
  0 
  0 
  0 
  0 
  0 
  0  \\ \hline 29&
  0 
  0 
  0 
  0 
  0 
  0 
  0 
  0 
  0 
  0  &
  0 
  0 
  0 
  0 
  0 
  0 
  0 
  0 
  0 
  0  &
  0 
  0 
  0 
  0 
  0 
  0 
  0 
  0 
  0 
  0  &
  0 
  0 
  0 
  0 
  0 
  0 
  0 
  0 
  0 
  0  \\ \hline 30&
  0 
  0 
  0 
  0 
  0 
  0 
  0 
  0 
  0 
  0  &
  0 
  0 
  0 
  0 
  0 
  0 
  0 
  0 
  0 
  0  &
  0 
  0 
  0 
  0 
  0 
  0 
  0 
  0 
  0 
  0  &
  0 
  0 
  0 
  0 
  0 
  0 
  0 
  0 
  0 
  0  \\ \hline 31&
  0 
  0 
  0 
  0 
  0 
  0 
  0 
  0 
  0 
  0  &
  0 
  0 
  0 
  0 
  0 
  0 
  0 
  0 
  0 
  0  &
  0 
  0 
  0 
  0 
  0 
  0 
  0 
  0 
  0 
  0  &
  0 
  0 
  0 
  0 
  0 
  0 
  0 
  0 
  0 
  0  \\ \hline 32&
  0 
  0 
  0 
  0 
  0 
  0 
  0 
  0 
  0 
  0  &
  0 
  0 
  0 
  0 
  0 
  0 
  0 
  0 
  0 
  0  &
  0 
  0 
  0 
  0 
  0 
  0 
  0 
  0 
  0 
  0  &
  0 
  0 
  0 
  0 
  0 
  0 
  0 
  0 
  0 
  0  \\ \hline 33&
  0 
  0 
  0 
  0 
  0 
  0 
  0 
  0 
  0 
  0  &
  0 
  0 
  0 
  0 
  0 
  0 
  0 
  0 
  0 
  0  &
  0 
  0 
  0 
  0 
  0 
  0 
  0 
  0 
  0 
  0  &
  0 
  0 
  0 
  0 
  0 
  0 
  0 
  0 
  0 
  0  \\ \hline 34&
  0 
  0 
  0 
  0 
  0 
  0 
  0 
  0 
  0 
  0  &
  0 
  0 
  0 
  0 
  0 
  0 
  0 
  0 
  0 
  0  &
  0 
  0 
  0 
  0 
  0 
  0 
  0 
  0 
  0 
  0  &
  0 
  0 
  0 
  0 
  0 
  0 
  0 
  0 
  0 
  0  \\ \hline 35&
  0 
  0 
  0 
  0 
  0 
  0 
  0 
  0 
  0 
  0  &
  0 
  0 
  0 
  0 
  0 
  0 
  0 
  0 
  0 
  0  &
  0 
  0 
  0 
  0 
  0 
  0 
  0 
  0 
  0 
  0  &
  0 
  0 
  0 
  0 
  0 
  0 
  0 
  0 
  0 
  0  \\ \hline 36&
  0 
  0 
  0 
  0 
  0 
  0 
  0 
  0 
  0 
  0  &
  0 
  0 
  0 
  0 
  0 
  0 
  0 
  0 
  0 
  0  &
  0 
  0 
  0 
  0 
  0 
  0 
  0 
  0 
  0 
  0  &
  0 
  0 
  0 
  0 
  0 
  0 
  0 
  0 
  0 
  0  \\ \hline 37&
  0 
  0 
  0 
  0 
  0 
  0 
  0 
  0 
  0 
  0  &
  0 
  0 
  0 
  0 
  0 
  0 
  0 
  0 
  0 
  0  &
  0 
  0 
  0 
  0 
  0 
  0 
  0 
  0 
  0 
  0  &
  0 
  0 
  0 
  0 
  0 
  0 
  0 
  0 
  0 
  0  \\ \hline
\end{tabular} \caption{Matrix A} \end{table}

\begin{table}\label{table:B} \begin{tabular}{|l|l|l|l|l|} \hline $B$&{\small  1 \hspace{4pt} $\cdots$\hspace{5pt} 5 \hspace{4pt}$\cdots$ \hspace{5pt}9} &{\small 10 \hspace{3pt}  $\cdots$ \hspace{3pt}15 \hspace{3pt}$\cdots$\hspace{3pt} 19 }& {\small 20 \hspace{4pt} $\cdots$ 25 \hspace{4pt} $\cdots$ 29 }& {\small 30 \hspace{11pt} $\cdots$ \hspace{11pt} 37 } \\ \hline
38& 1  1    0 
  0 
  0 
  0 
  0 
  0 
  0  &
  0 
  0 
  0 
  0 
  0 
  0 
  0 
  0 
  0 
  0  &
  0 
  0 
  0 
  0 
  0 
  0 
  0 
  0   1
  0  &
  0 
  0 
  0 
  0 
  0 
  0   1
  0  \\ \hline 39& 1  1
  0 
  0 
  0 
  0 
  0 
  0 
  0  &
  0 
  0 
  0 
  0 
  0 
  0   1
  0 
  0 
  0  &
  0 
  0 
  0 
  0 
  0 
  0   1
  0   1
  0  &
  0 
  0 
  0 
  0   1
  0 
  0 
  0  \\ \hline 40& 1
  0 
  0 
  0 
  0 
  0 
  0 
  0 
  0  &
  0 
  0 
  0 
  0 
  0 
  0 
  0 
  0 
  0 
  0  &
  0 
  0 
  0 
  0 
  0 
  0 
  0 
  0 
  0 
  0  &
  0 
  0 
  0 
  0   1
  0 
  0 
  0  \\ \hline 41& 1
  0 
  0 
  0 
  0 
  0 
  0 
  0 
  0  &
  0 
  0 
  0 
  0 
  0 
  0 
  0 
  0 
  0 
  0  &
  0 
  0 
  0 
  0 
  0 
  0 
  0 
  0 
  0 
  0  &
  0 
  0 
  0 
  0 
  0 
  0 
  0 
  0  \\ \hline 42& 1
  0 
  0 
  0 
  0 
  0 
  0 
  0 
  0  &
  0 
  0 
  0 
  0 
  0 
  0 
  0 
  0 
  0 
  0  &
  0 
  0 
  0 
  0 
  0 
  0 
  0 
  0 
  0 
  0  &
  0 
  0 
  0 
  0   1
  0 
  0 
  0  \\ \hline 43&
  0 
  0 
  0 
  0 
  0 
  0 
  0 
  0 
  0  &
  0 
  0 
  0 
  0 
  0 
  0 
  0 
  0 
  0 
  0  &
  0 
  0 
  0 
  0 
  0 
  0 
  0 
  0 
  0 
  0  &
  0 
  0 
  0 
  0 
  0 
  0 
  0 
  0  \\ \hline 44&
  0 
  0   1
  0 
  0 
  0 
  0 
  0 
  0  &
  0 
  0 
  0 
  0 
  0 
  0 
  0 
  0 
  0 
  0  &
  0 
  0 
  0 
  0 
  0 
  0 
  0 
  0 
  0 
  0  &
  0 
  0 
  0 
  0 
  0 
  0 
  0 
  0  \\ \hline 45& 1  1
  0 
  0 
  0 
  0 
  0 
  0 
  0  &
  0 
  0   1
  0 
  0 
  0 
  0 
  0 
  0 
  0  &
  0 
  0 
  0 
  0 
  0 
  0   1
  0 
  0 
  0  &
  0 
  0 
  0 
  0   1
  0 
  0 
  0  \\ \hline 46& 1
  0 
  0 
  0 
  0 
  0 
  0 
  0 
  0  &
  0 
  0 
  0 
  0 
  0 
  0 
  0 
  0 
  0 
  0  &
  0 
  0 
  0 
  0 
  0 
  0 
  0 
  0 
  0 
  0  &
  0 
  0 
  0 
  0 
  0 
  0 
  0 
  0  \\ \hline 47&
  0 
  0 
  0 
  0 
  0 
  0 
  0 
  0 
  0  &
  0 
  0 
  0 
  0 
  0 
  0 
  0 
  0 
  0 
  0  &
  0 
  0 
  0 
  0 
  0 
  0 
  0 
  0 
  0 
  0  &
  0 
  0 
  0 
  0 
  0 
  0 
  0 
  0  \\ \hline 48&
  0 
  0 
  0 
  0 
  0 
  0 
  0 
  0 
  0  &
  0 
  0 
  0 
  0 
  0 
  0 
  0 
  0 
  0 
  0  &
  0 
  0 
  0 
  0 
  0 
  0 
  0 
  0 
  0 
  0  &
  0 
  0 
  0 
  0 
  0 
  0 
  0 
  0  \\ \hline 49&
  0 
  0 
  0 
  0 
  0 
  0 
  0 
  0 
  0  &
  0 
  0 
  0 
  0 
  0 
  0 
  0 
  0 
  0 
  0  &
  0 
  0 
  0 
  0 
  0 
  0 
  0 
  0 
  0 
  0  &
  0 
  0 
  0 
  0 
  0 
  0 
  0 
  0  \\ \hline 50&
  0 
  0 
  0 
  0 
  0 
  0 
  0 
  0 
  0  &
  0 
  0 
  0 
  0 
  0 
  0 
  0 
  0 
  0 
  0  &
  0 
  0 
  0 
  0 
  0 
  0 
  0 
  0 
  0 
  0  &
  0 
  0 
  0 
  0 
  0 
  0 
  0 
  0  \\ \hline 51&
  0   1
  0 
  0 
  0 
  0 
  0 
  0 
  0  &
  0 
  0 
  0 
  0 
  0 
  0 
  0 
  0 
  0 
  0  &
  0 
  0 
  0 
  0 
  0 
  0 
  0 
  0 
  0 
  0  &
  0 
  0 
  0 
  0 
  0 
  0 
  0 
  0  \\ \hline 52&
  0 
  0 
  0 
  0 
  0 
  0 
  0 
  0 
  0  &
  0 
  0 
  0 
  0 
  0 
  0 
  0 
  0 
  0 
  0  &
  0 
  0 
  0 
  0 
  0 
  0 
  0 
  0 
  0 
  0  &
  0 
  0 
  0 
  0 
  0 
  0 
  0 
  0  \\ \hline 53&
  0 
  0   1
  0 
  0 
  0 
  0 
  0 
  0  &
  0 
  0 
  0 
  0 
  0 
  0 
  0 
  0 
  0 
  0  &
  0 
  0 
  0 
  0 
  0 
  0 
  0 
  0 
  0 
  0  &
  0 
  0 
  0 
  0 
  0 
  0 
  0 
  0  \\ \hline 54&
  0 
  0 
  0 
  0 
  0 
  0 
  0 
  0 
  0  &
  0 
  0 
  0 
  0 
  0 
  0 
  0 
  0 
  0 
  0  &
  0 
  0 
  0 
  0 
  0 
  0 
  0 
  0 
  0 
  0  &
  0 
  0 
  0 
  0 
  0 
  0 
  0 
  0  \\ \hline 55&
  0 
  0 
  0 
  0 
  0 
  0 
  0 
  0 
  0  &
  0 
  0 
  0 
  0 
  0 
  0 
  0 
  0 
  0 
  0  &
  0 
  0 
  0 
  0 
  0 
  0 
  0 
  0 
  0 
  0  &
  0 
  0 
  0 
  0 
  0 
  0 
  0 
  0  \\ \hline 56&
  0 
  0 
  0 
  0 
  0 
  0 
  0 
  0 
  0  &
  0 
  0 
  0 
  0 
  0 
  0 
  0 
  0 
  0 
  0  &
  0 
  0 
  0 
  0 
  0 
  0 
  0 
  0 
  0 
  0  &
  0 
  0 
  0 
  0 
  0 
  0 
  0 
  0  \\ \hline 57&
  0 
  0 
  0 
  0 
  0 
  0 
  0 
  0 
  0  &
  0 
  0 
  0 
  0 
  0 
  0 
  0 
  0 
  0 
  0  &
  0 
  0 
  0 
  0 
  0 
  0 
  0 
  0 
  0 
  0  &
  0 
  0 
  0 
  0 
  0 
  0 
  0 
  0  \\ \hline 58&
  0 
  0 
  0 
  0 
  0 
  0 
  0 
  0 
  0  &
  0 
  0 
  0 
  0 
  0 
  0 
  0 
  0 
  0 
  0  &
  0 
  0 
  0 
  0  
  0 
  0 
  0 
  0 
  0 
  0  &
  0 
  0 
  0 
  0 
  0 
  0 
  0 
  0  \\ \hline 59&
  0 
  0 
  0 
  0 
  0 
  0 
  0 
  0 
  0  &
  0 
  0 
  0 
  0 
  0 
  0 
  0 
  0 
  0 
  0  &
  0 
  0 
  0 
  0 
  0 
  0 
  0 
  0 
  0 
  0  &
  0 
  0 
  0 
  0 
  0 
  0 
  0 
  0  \\ \hline 60&
  0 
  0 
  0 
  0 
  0 
  0 
  0 
  0 
  0  &
  0 
  0 
  0 
  0 
  0 
  0 
  0 
  0 
  0 
  0  &
  0 
  0 
  0 
  0 
  0 
  0 
  0 
  0 
  0 
  0  &
  0 
  0 
  0 
  0 
  0 
  0 
  0 
  0  \\ \hline 61&
  0 
  0 
  0 
  0 
  0 
  0 
  0 
  0 
  0  &
  0 
  0 
  0 
  0 
  0 
  0 
  0 
  0 
  0 
  0  &
  0 
  0 
  0 
  0 
  0 
  0 
  0 
  0 
  0 
  0  &
  0 
  0 
  0 
  0 
  0 
  0 
  0 
  0  \\ \hline 62&
  0 
  0 
  0 
  0 
  0 
  0 
  0 
  0 
  0  &
  0 
  0 
  0 
  0 
  0 
  0 
  0 
  0 
  0 
  0  &
  0 
  0 
  0 
  0 
  0 
  0 
  0 
  0 
  0 
  0  &
  0 
  0 
  0 
  0 
  0 
  0 
  0 
  0  \\ \hline 63&
  0 
  0 
  0 
  0 
  0 
  0 
  0 
  0 
  0  &
  0 
  0 
  0 
  0 
  0 
  0 
  0 
  0 
  0 
  0  &
  0 
  0 
  0 
  0 
  0 
  0 
  0 
  0 
  0 
  0  &
  0 
  0 
  0 
  0 
  0 
  0 
  0 
  0  \\ \hline 64&
  0 
  0 
  0 
  0 
  0 
  0 
  0 
  0 
  0  &
  0 
  0 
  0 
  0 
  0 
  0 
  0 
  0 
  0 
  0  &
  0 
  0 
  0 
  0 
  0 
  0 
  0 
  0 
  0 
  0  &
  0 
  0 
  0 
  0 
  0 
  0 
  0 
  0  \\ \hline 65&
  0 
  0 
  0 
  0 
  0 
  0 
  0 
  0 
  0  &
  0 
  0 
  0 
  0 
  0 
  0 
  0 
  0 
  0 
  0  &
  0 
  0 
  0 
  0 
  0 
  0 
  0 
  0   1
  0  &
  0 
  0 
  0 
  0 
  0 
  0 
  0 
  0  \\ \hline 66&
  0 
  0 
  0 
  0 
  0 
  0 
  0 
  0 
  0  &
  0 
  0 
  0 
  0 
  0 
  0 
  0 
  0 
  0 
  0  &
  0 
  0 
  0 
  0 
  0 
  0 
  0 
  0 
  0 
  0  &
  0 
  0 
  0 
  0 
  0 
  0 
  0 
  0  \\ \hline 67&
  0 
  0 
  0 
  0 
  0 
  0 
  0 
  0 
  0  &
  0 
  0 
  0 
  0 
  0 
  0 
  0 
  0 
  0 
  0  &
  0 
  0 
  0 
  0 
  0 
  0 
  0 
  0 
  0 
  0  &
  0 
  0 
  0 
  0 
  0 
  0 
  0 
  0  \\ \hline 68&
  0 
  0 
  0 
  0 
  0 
  0 
  0 
  0 
  0  &
  0 
  0 
  0 
  0 
  0 
  0 
  0 
  0 
  0 
  0  &
  0 
  0 
  0 
  0 
  0 
  0 
  0 
  0 
  0 
  0  &
  0 
  0 
  0 
  0 
  0 
  0 
  0 
  0  \\ \hline 69&
  0 
  0 
  0 
  0 
  0 
  0 
  0 
  0 
  0  &
  0 
  0 
  0 
  0 
  0 
  0 
  0 
  0 
  0 
  0  &
  0 
  0 
  0 
  0 
  0 
  0 
  0 
  0 
  0 
  0  &
  0 
  0 
  0 
  0 
  0 
  0 
  0 
  0  \\ \hline 70&
  0 
  0 
  0 
  0 
  0 
  0 
  0 
  0 
  0  &
  0 
  0 
  0 
  0 
  0 
  0 
  0 
  0 
  0 
  0  &
  0 
  0 
  0 
  0 
  0 
  0 
  0 
  0 
  0 
  0  &
  0 
  0 
  0 
  0 
  0 
  0 
  0 
  0  \\ \hline 71&
  0 
  0 
  0 
  0 
  0 
  0 
  0 
  0 
  0  &
  0 
  0 
  0 
  0 
  0 
  0 
  0 
  0 
  0 
  0  &
  0 
  0 
  0 
  0 
  0 
  0 
  0 
  0 
  0 
  0  &
  0 
  0 
  0 
  0 
  0 
  0 
  0 
  0  \\ \hline 72&
  0 
  0 
  0 
  0 
  0 
  0 
  0 
  0 
  0  &
  0 
  0 
  0 
  0 
  0 
  0 
  0 
  0 
  0 
  0  &
  0 
  0 
  0 
  0 
  0 
  0 
  0 
  0 
  0 
  0  &
  0 
  0 
  0 
  0   1
  0 
  0 
  0  \\ \hline 73&
  0 
  0 
  0 
  0 
  0 
  0 
  0 
  0 
  0  &
  0 
  0 
  0 
  0 
  0 
  0 
  0 
  0 
  0 
  0  &
  0 
  0 
  0 
  0 
  0 
  0 
  0 
  0 
  0 
  0  &
  0 
  0 
  0 
  0 
  0 
  0 
  0 
  0  \\ \hline 74&
  0 
  0 
  0 
  0 
  0 
  0 
  0 
  0 
  0  &
  0 
  0 
  0 
  0 
  0 
  0 
  0 
  0 
  0 
  0  &
  0 
  0 
  0 
  0 
  0 
  0 
  0 
  0 
  0 
  0  &
  0 
  0 
  0 
  0 
  0 
  0 
  0 
  0  \\ \hline 75&
  0 
  0 
  0 
  0 
  0 
  0 
  0 
  0 
  0  &
  0 
  0 
  0 
  0 
  0 
  0 
  0 
  0 
  0 
  0  &
  0 
  0 
  0 
  0 
  0 
  0 
  0 
  0 
  0 
  0  &
  0 
  0 
  0 
  0 
  0 
  0 
  0 
  0  \\ \hline 76& 1  1  1
  0 
  0   1  1  1
  0  &
  0 
  0 
  0 
  0 
  0 
  0 
  0 
  0   1
  0  &
  0 
  0 
  0 
  0 
  0 
  0 
  0 
  0 
  0 
  0  &
  0 
  0 
  0 
  0 
  0 
  0 
  0 
  0  \\ \hline 77&
  0 
  0 
  0 
  0 
  0 
  0 
  0 
  0 
  0  &
  0 
  0 
  0 
  0 
  0 
  0 
  0 
  0 
  0 
  0  &
  0 
  0 
  0 
  0 
  0 
  0 
  0 
  0 
  0 
  0  &
  0 
  0 
  0 
  0 
  0 
  0 
  0 
  0  \\ \hline
\end{tabular} \caption{Matrix B}\end{table}

\end{document}